\documentclass[twocolumn]{aastex631}
\usepackage{verbatim}
\usepackage[utf8]{inputenc}
\usepackage{cellspace}
\usepackage{amssymb}
\usepackage{amsmath}
\usepackage{multirow}

\newcommand{\msun}{$\rm M_\odot$}
\newcommand{\fesc}{$f_{\rm{esc, 912}}$}
\newcommand{\kms}{\rm km\ s^{-1}}
\newcommand{\ergs}{\rm erg\ s^{-1}}
\newcommand{\lya}{Ly$\alpha$} 
\graphicspath{ {figures/} }

\begin{document}

\title{Haro 11: The Spatially Resolved Lyman Continuum Sources}

\correspondingauthor{Lena Komarova}
\email{komarova@umich.edu}

\author[0000-0002-5235-7971]{Lena Komarova}
\affiliation{Astronomy Department, University of Michigan, 1085 South University Avenue, Ann Arbor, MI 48109, USA}

\author[0000-0002-5808-1320]{M. S. Oey}
\affiliation{Astronomy Department, University of Michigan, 1085 South University Avenue, Ann Arbor, MI 48109, USA}

\author[0000-0003-4857-8699]{Svea Hernandez}
\affiliation{AURA for ESA, Space Telescope Science Institute, 3700 San Martin Drive, Baltimore, MD 21218, USA}

\author[0000-0002-8192-8091]{Angela Adamo}
\affiliation{Department of Astronomy, Oskar Klein Centre, Stockholm University, AlbaNova University Centre, SE-106 91 Stockholm, Sweden}

\author[0000-0002-0923-8352]{Mattia Sirressi}
\affiliation{Department of Astronomy, Oskar Klein Centre, Stockholm University, AlbaNova University Centre, SE-106 91 Stockholm, Sweden}

\author[0000-0003-2685-4488]{Claus Leitherer}
\affiliation{AURA for ESA, Space Telescope Science Institute, 3700 San Martin Drive, Baltimore, MD 21218, USA}

\author[0000-0002-8823-9723]{J. M. Mas-Hesse}
\affiliation{Centro de Astrobiología – Dept. de Astrofísica (CSIC-INTA), 28850 Madrid, Spain}

\author[0000-0002-3005-1349]{Göran Östlin}
\affiliation{Department of Astronomy, Oskar Klein Centre, Stockholm University, AlbaNova University Centre, SE-106 91 Stockholm, Sweden}

\author[0000-0002-2397-206X]{Edmund Hodges-Kluck}
\affiliation{NASA Goddard Space Flight Center, Greenbelt, MD 20771, USA}

\author[0000-0001-8068-0891]{Arjan Bik}
\affiliation{Department of Astronomy, Oskar Klein Centre, Stockholm University, AlbaNova University Centre, SE-106 91 Stockholm, Sweden}

\author[0000-0001-8587-218X]{Matthew J. Hayes}
\affiliation{Department of Astronomy, Oskar Klein Centre, Stockholm University, AlbaNova University Centre, SE-106 91 Stockholm, Sweden}

\author[0000-0002-6790-5125]{Anne E. Jaskot}
\affiliation{Astronomy Department, Williams College, Williamstown,
MA 01267, USA}

\author[0000-0002-3635-0112]{Daniel Kunth}
\affiliation{Institut d'Astrophysique, Paris, 98 bis Boulevard Arago, F-75014 Paris, France}

\author[0000-0003-4207-0245]{Peter Laursen}
\affiliation{Cosmic Dawn Center (DAWN)}
\affiliation{Niels Bohr Institute, University of Copenhagen, Jagtvej 128, DK-2200, Copenhagen N, Denmark}

\author[0000-0003-0470-8754]{Jens Melinder}
\affiliation{Department of Astronomy, Oskar Klein Centre, Stockholm University, AlbaNova University Centre, SE-106 91 Stockholm, Sweden}

\author[0000-0002-9204-3256]{T. Emil Rivera-Thorsen}
\affiliation{Department of Astronomy, Oskar Klein Centre, Stockholm University, AlbaNova University Centre, SE-106 91 Stockholm, Sweden}

\accepted{March 31, 2024}

\shorttitle{The LyC Source in Haro 11}
\shortauthors{Komarova et al.}

\begin{abstract}
As the nearest confirmed Lyman continuum (LyC) emitter, Haro 11 is an exceptional laboratory for studying LyC escape processes crucial to cosmic reionization. Our new HST/COS G130M/1055 observations of its three star-forming knots now reveal that the observed LyC originates in Knots~B and C, with $903 - 912$ \AA\ luminosities of $1.9\pm1.5 \times 10^{40}~\rm erg~s^{-1}$ and $0.9\pm0.7 \times 10^{40}~\rm erg~s^{-1}$, respectively. We derive local escape fractions \fesc~$= 3.4\pm2.9$\% and $5.1\pm4.3$\% for Knots B and C, respectively. Our Starburst99 modeling shows dominant populations on the order of $\sim1-4$~Myr and $1-2\times10^7$~\msun~in each knot, with the youngest population in Knot B. Thus, the knot with the strongest LyC detection has the highest LyC production. However, LyC escape is likely less efficient in Knot B than in Knot C due to higher neutral gas covering. Our results therefore stress the importance of the intrinsic ionizing luminosity, and not just the escape fraction, for LyC detection. Similarly, the \lya\ escape fraction does not consistently correlate with LyC flux, nor do narrow \lya\ red peaks. High observed \lya~luminosity and low \lya\ peak velocity separation, however, do correlate with higher LyC escape. Another insight comes from the undetected Knot A, which drives the Green Pea properties of Haro 11. Its density-bounded conditions suggest highly anisotropic LyC escape. Finally, both of the LyC-leaking Knots, B and C, host ultra-luminous X-ray sources (ULXs). While stars strongly dominate over the ULXs in LyC emission, this intriguing coincidence underscores the importance of unveiling the role of accretors in LyC escape and reionization.
 
\end{abstract}


\section{Introduction}
\label{sec:intro}

The ionizing sources and physical mechanisms responsible for cosmic reionization at $\rm z>6$ remain a critical unsolved problem in cosmology. The major contenders for providing the required LyC are active galactic nuclei (AGN) and massive stars in starbursts, with their relative contributions to reionization still uncertain. Accreting sources other than AGN, e.g. ultra luminous X-ray sources (ULXs), may also play a significant role in ionizing the IGM \citep{Madau2017, Ross2017, Sazonov2018}. Some studies show that AGN could produce sufficient energy to reionize the universe \cite[e.g.,][]{Madau2015, Giallongo2015}; while others suggest that the AGN number density and ionizing emissivity were too low in the early universe \citep{Shankar2007, Fontanot2012, Hassan2018, Faucher2020}, pointing to star-forming galaxies as the dominant source. On the other hand, while dwarf starbursts seem to be promising candidates for reionization agents \citep{Bouwens2012, Sharma2017, Yeh2023}, they may not have sufficiently high escape fractions  \cite[e.g.,][]{Fontanot2014}. JWST is now revealing the blue UV slopes of the earliest, $z = 8-16$, galaxies, pointing to young and dust-poor stellar populations \citep{Cullen2023, Cullen2023dust, Topping2023, Morales2023}, in further support of galaxy-driven reionization. JWST is moreover uncovering an abundance of high-redshift starbursts with high ionizing photon production efficiencies, implying that galaxies could have reionized the universe with somewhat lower escape fractions than previously assumed \citep{ Matthee2023, Atek2024}.

Knowing the relevant UV sources is only half of the problem, however. The other key question is how LyC escapes the immediate environment of the source without being absorbed by the local interstellar medium (ISM). The standard paradigm for stellar-driven reionization is that supernovae and stellar winds clear pathways in the ISM that become optically thin to LyC \cite[e.g.,][]{Clarke2002, Fujita2003, Ma2016}. There is now evidence of a radiation-driven feedback mode that also may enable LyC escape in the most extreme metal-poor starbursts such as extreme Green Peas (GPs) \citep{Jaskot2017, Komarova2021, Flury2022a, Flury2022b}. Similarly, accretion-driven sources such as AGN or X-ray binaries may create optically thin channels through winds and jets \citep{Smith2020}.  

\begin{figure}
\includegraphics[width=\columnwidth]{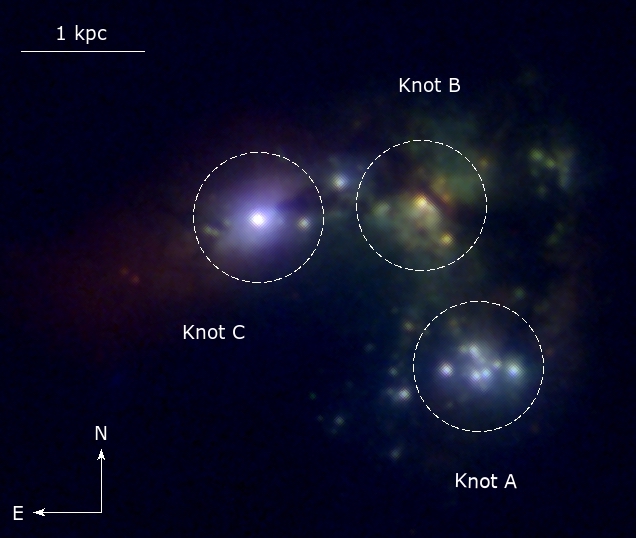}
\caption{HST color composite image of Haro 11, indicating Knots A, B, and C, as well as the COS apertures (red = 5500~\AA~continuum, F550M; green = 3360~\AA~continuum, F336W; blue = Ly$\alpha$, F122M).}
\label{fig:Fig1}
\end{figure}

Haro 11 is an extreme dwarf starburst galaxy with dozens of young massive clusters \citep{Adamo2010, Sirressi2022} and a gas consumption timescale of 50~Myr \citep{Ostlin2021}. It is one of the most important local ($z = 0.021$) galaxies for advancing our understanding of cosmic reionization, and a wealth of data across the electromagnetic spectrum exist for this object. Haro 11 is the first local Lyman continuum emitter (LCE) to be observationally confirmed \citep{Bergvall2006, Leitet}, and is moreover the closest one, lying at a distance of only 88.5 Mpc \citep[assuming $H_0 = 73~\rm km~s^{-1}~Mpc^{-1}$;][hereafter S22]{Sirressi2022}.  With its intense star formation triggered by a dwarf galaxy merger \citep[e.g.,][]{Ostlin2001, Ostlin2015}, Haro 11 is dominated by three starburst knots: A, B, and C \citep{Kunth2003}, indicated in Figure \ref{fig:Fig1}. While Knot A is a purely star-forming region, with properties consistent with those of Green Peas \citep{Keenan2017}, Knots B and C host ULXs \citep{Prestwich2015}, including a possible low-luminosity AGN (LLAGN) \citep{Gross2021}. Thus, Haro 11 is a prime laboratory for pinpointing the nature of UV sources and LyC escape processes crucial to reionization. 

However, the initial LyC detection \citep{Bergvall2006} did not resolve which of the three knots is/are responsible for the LyC emission, since the $30\arcsec\times30\arcsec$ FUSE aperture encompassed the entirety of the galaxy. Unveiling the exact source(s) of the Haro 11 LyC leakage is necesssary to clarify the relative roles of compact objects and massive stars in this prototypical object. The three knots vary drastically in Ly$\alpha$ emission properties, extinction, stellar populations, and gas properties. So, putting the LyC detection in context of knot properties reveals the connection between environment and LyC escape.
In this paper, we present HST/COS observations of the three knots at wavelengths below the Lyman limit, revealing the LyC-emitting sources. Section \ref{obs} contains the details of observations and data analysis. We present our results in Section \ref{sec:Results}, discuss cosmological implications in Section \ref{sec:Discussion}, and summarize our main conclusions in Section \ref{sec:Conclusions}. 

\section{Observations and Analysis}
\label{obs}
We obtained HST/COS FUV spectra of Knots A, B, and C in Haro 11 (Cycle 28, program ID 16260, PI: Oey). We used the medium-resolution G130M grating in TIME-TAG mode, coupled with the $2.5\arcsec$ Primary Science Aperture (PSA), with all four grating offset positions (FP-POS = ALL).  The observations were taken with the grating centered on 1055~\AA, and resolving power $R \sim 10,000$.
A total of 13 Visits over the course of 24 orbits were obtained, in 2021 January, May, October, November, and December; and 2022 May and July. For each Visit on each knot, we obtained 4 sub-exposures at different focal plane offset positions in the wavelength range 900 -- 1200~\AA.  Some sub-exposures had unusable data due to acquisition failures and resulted in additional Visits. All sub-exposures used for the LyC measurements are listed in Table \ref{table:observations}.

The raw COS observations were calibrated with \texttt{CalCOS} version 3.4.0 \citep{Soderblom21}. Given that the observations for the individual knots were obtained throughout several different Visits, with more than one association file, we combined their individual calibrated \texttt{x1d} files using the \texttt{IDL} software by \citet{Danforth2010,Danforth2016}. We combined the spectra, weighting by exposure times and taking into account the data quality (DQ) flags set by the standard \texttt{CalCOS} pipeline.

A critical step in the calibration of Lyman continuum observations is an accurate background correction. The standard \texttt{CalCOS} pipeline estimates the background contribution in the region of the detector where the science spectrum is located, by computing the average counts in two predefined regions external to the science extraction region. For the COS FUV detector, these two predefined regions are typically located below and above the science extraction box, depending on the lifetime position (LP) used at the time of the observations. The COS team has reported that the background levels are correlated with solar activity \citep{Dashtamirova2019}, detector gain and high voltage (HV). Additionally, they have found that the background levels vary throughout the detector and with time, with slightly higher levels towards the edge of the detector. 

\begin{deluxetable}{ccc}
\tablecolumns{3}
\tablecaption{COS G130M/1055 Observations
\label{table:observations}}
\tablehead{
\colhead{Target} & \colhead{UT Start Date} & \colhead{Exposure time \tablenotemark{\rm \scriptsize a} (s)}}
\startdata
 & 2021-05-06 & 2,672 \\ 
& 2021-05-07  & 3,406\\ 
Knot A & 2021-10-13 & 2,306 \\
& 2021-12-01 & 2,306 \\
 & 2021-12-01 & 3,862 \\
 & 2022-05-20 & 2,306\\ 
 & &Total: 16,858 \\
 \hline
 & 2022-05-20 & 4,924 \\ 
Knot B &  2022-07-05 & 4,712 \\
& 2022-07-07 & 4,713 \\
& & Total: 14,349 \\
\hline
 & 2021-01-27 & 4,841 \\
 Knot C & 2021-05-12 & 3,500 \\
  &  2021-01-25 & 4,847 \\
  &  2021-12-01 & 2,306 \\
  & & Total: 11,994
\enddata 
\textbf{Notes.}
\tablenotetext{\rm a}{Exposure times used in all Visits incorporated in our LyC measurements.}
\end{deluxetable}

To investigate whether the background correction was accurately accounting for the expected detector contribution in the science extraction regions, we collapsed the 2-dimensional \texttt{flt} images in the dispersion direction. 
For the G130M/1055 configuration observed at Lifetime Position 2, the background region boundaries are located at pixels $y\sim448$ and $y\sim728$, 
with predefined widths of 51 pixels. Inspection of the collapsed profiles showed that in all exposures the background region centered on pixel $y\sim728$ showed a slightly higher background level towards the edge of the detector than that observed closer to the science extraction region. This in turn caused a slight oversubtraction of the science spectra. To improve the background correction, we modified the location of the background regions in the extraction tab reference file (\texttt{XTRACTAB}) to be closer to the science extraction region (centered at $y\sim$ 588), centering the predefined regions around pixels $y\sim$ 505 and $y\sim$ 663. The background estimates using these new regions were $\sim3-6$\% lower than those calculated at the original background locations. 

Given the extended nature of the Haro 11 targets, we adopted a \texttt{BOXCAR} extraction technique, available in \texttt{CalCOS}. As detailed in \citet{James2022}, for extended targets, a broader extraction box may be necessary to collect the full extent of their flux. To select an optimal extraction height value for the science region, we explored varying the nominal size by $\pm$8 pixels, in steps of 2 pixels, aiming to improve the signal-to-noise of the extracted spectrum. We adopted extraction heights for the science regions in COS segment B of 63 (standard), 71, and 71 pixels for Knots A, B, and C, respectively.

We sum the high-quality spectra from each Visit on each knot, with total effective exposure times for Knots A, B, and C of 4.7, 4.0, and 3.3 hours, respectively (Table \ref{table:observations}). To maximize the signal-to-noise ratio, we moreover bin the spectra by 6 G130M/1055 resolution elements, and thus 60 pixels $\approx 0.6$~\AA.

\begin{figure}
\includegraphics[width=\columnwidth]{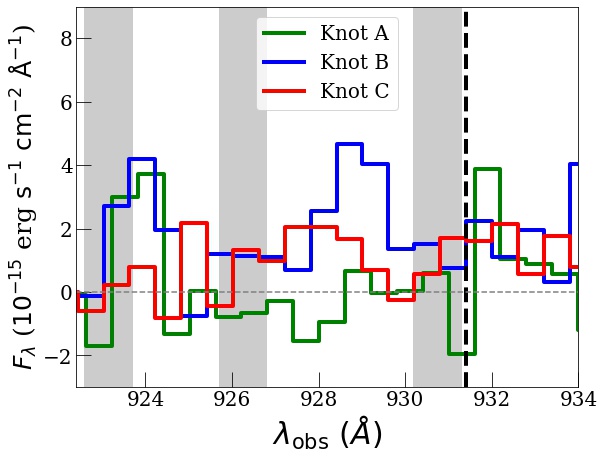}  
\caption{New HST/COS Lyman continuum observations of Knots A, B, and C in Haro 11. We measure the LyC in a 9~\AA~window between 922.4~\AA~(the left edge of the figure) and the redshifted Lyman limit at 931.4~\AA, shown as a black dashed line. The gray regions represent intervals that were excluded from LyC measurements due to geocoronal emission lines.}
\label{fig:lyc}
\end{figure}

Our reduced, median-combined COS LyC spectra for the three knots are shown in Figure \ref{fig:lyc}. The LyC region we observe is a 9 \AA~window below the redshifted Haro 11 Lyman limit of 931.4 \AA, as indicated in Figure \ref{fig:lyc}.  We compute the mean LyC flux densities $F_{912}$ in the three knots, excluding 1.1 \AA-wide geocoronal H Lyman series line regions  at 930.75 \AA, 923.15 \AA, and 926.25 \AA. The flux measurement remains the same within the error, when including these regions. 
There are no other known geocoronal features within the measured region. 
Our observed LyC fluxes and luminosities are shown in Table \ref{Table2}. We detect the strongest LyC emission in Knot B, which has about 2/3 of the total flux, leaving 1/3 in Knot C and none in Knot A. Our combined LyC flux density from Knots B and C is $F_{912} = (3.5\pm 2)\times10^{-15}~\rm erg~s^{-1}~cm^{-2}$~\AA$^{-1}$, consistent with the $(4.0\pm 0.9)\times10^{-15}~\rm erg~s^{-1}~cm^{-2}$~\AA$^{-1}$ measured by \cite{Leitet} in the $30\arcsec \times 30\arcsec$ FUSE aperture.

\begin{figure*}
\centering
\gridline{
    \fig{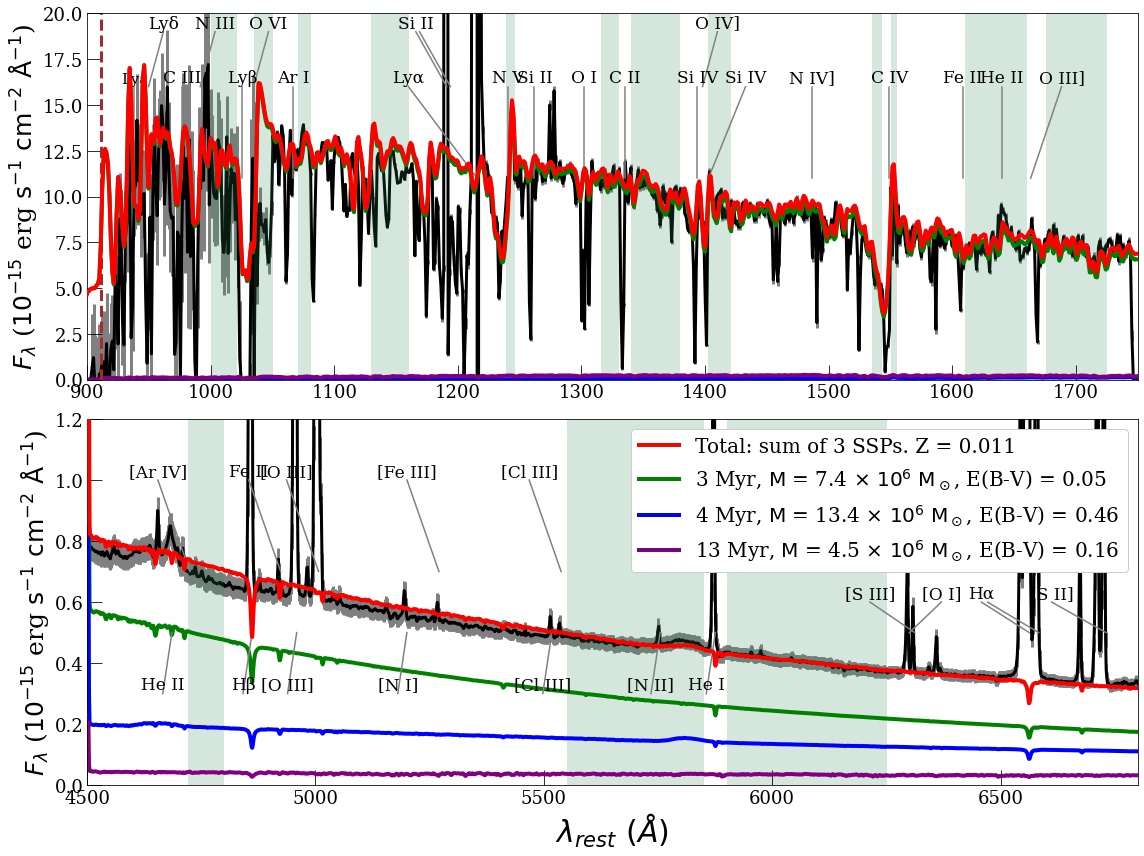}{0.9\textwidth}{}}
\vspace{-15pt}
\gridline{
    \fig{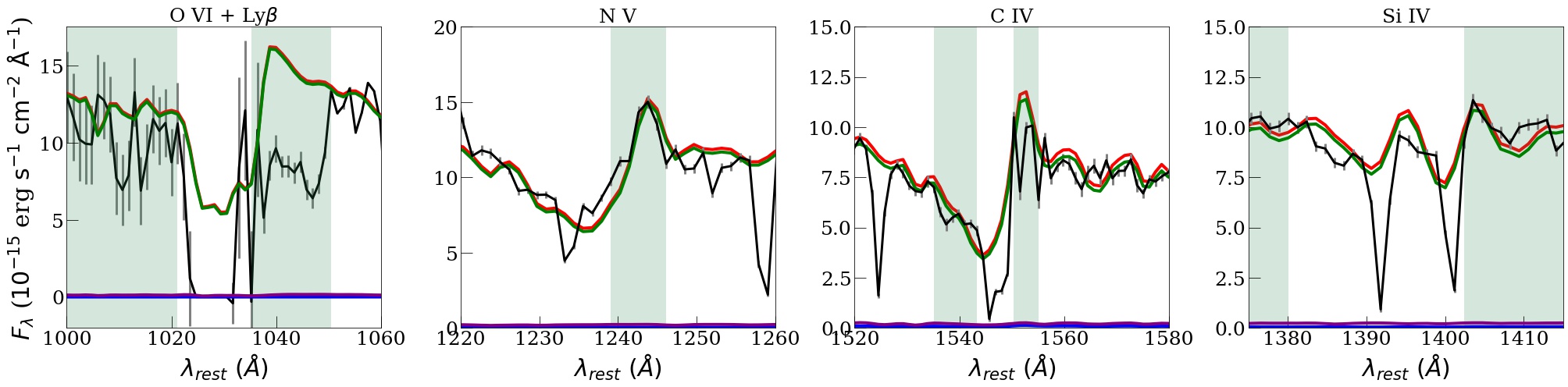}{0.9\textwidth}{}}
\caption{Top: Combined new COS G130M/1055 + \cite{Sirressi2022} G130M/1300 + G160M/1600 + MUSE  observations (black) of Haro 11 Knot A, and our model consisting of 3 SSPs, indicated by the colored lines as shown. The light green regions represent intervals that were used for fitting the models to the data. The rest was excluded to mask ISM and nebular lines, geocoronal emission, and detector gaps. Bottom: Zoom of age-sensitive O~VI, N~V, C~IV, and Si~IV P-Cygni profile fits.}
\label{fig:Astellarpops}
\end{figure*}

\begin{figure*}
\centering
\gridline{
    \fig{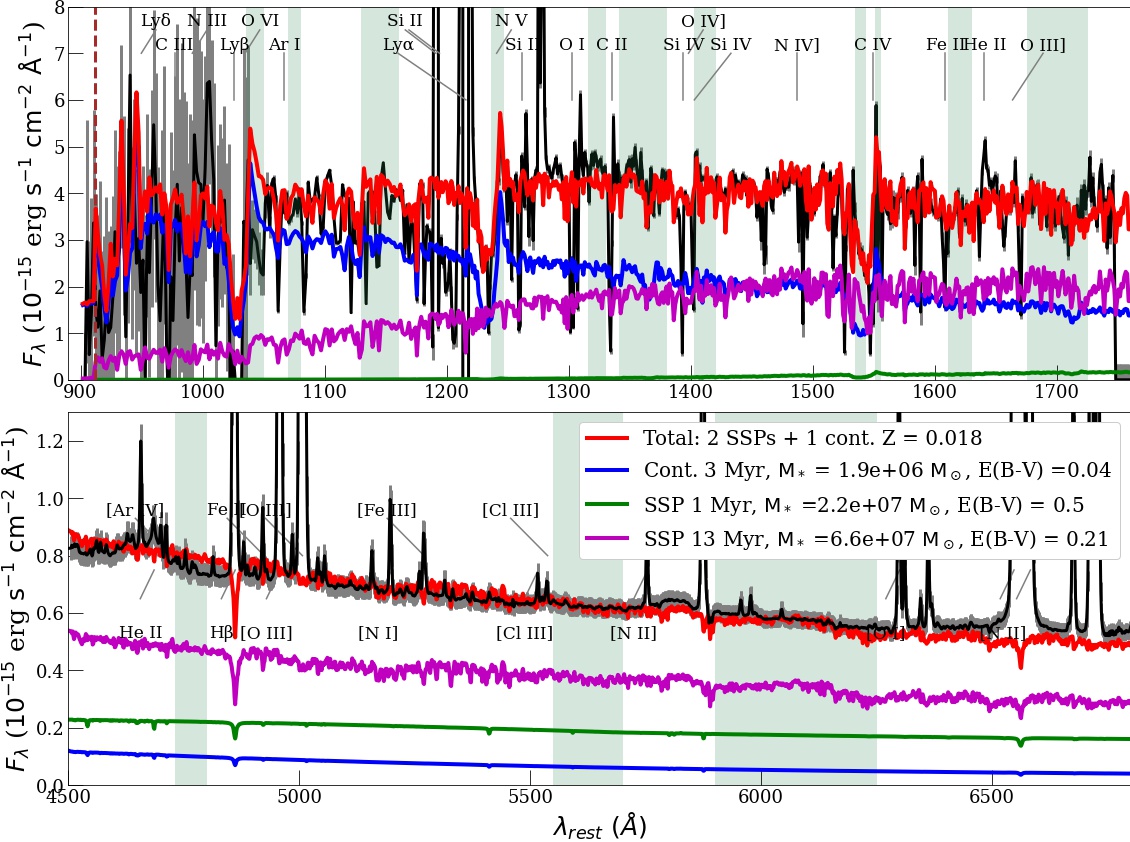}{0.9\textwidth}{}}
\vspace{-15pt}
\gridline{
    \fig{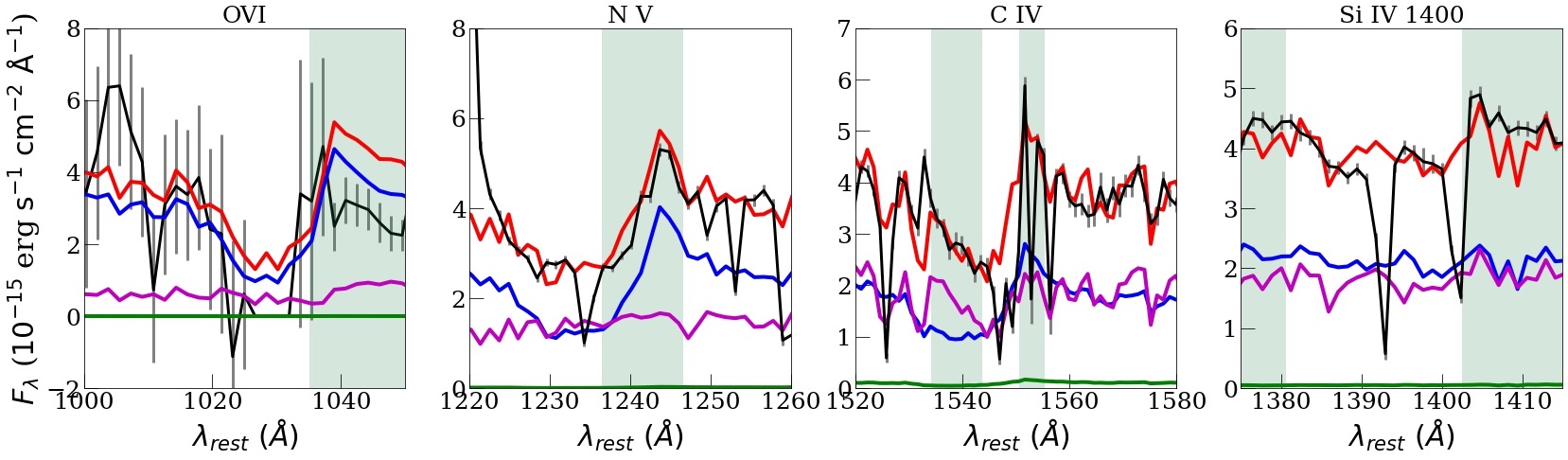}{0.9\textwidth}{}}
\caption{Same as Figure \ref{fig:Astellarpops} but for Knot B. The model assumes two single stellar populations and an episode of continuous star formation.}
\label{fig:Bstellarpops}
\end{figure*}

% KNOT C
\begin{figure*}
\centering
\gridline{
    \fig{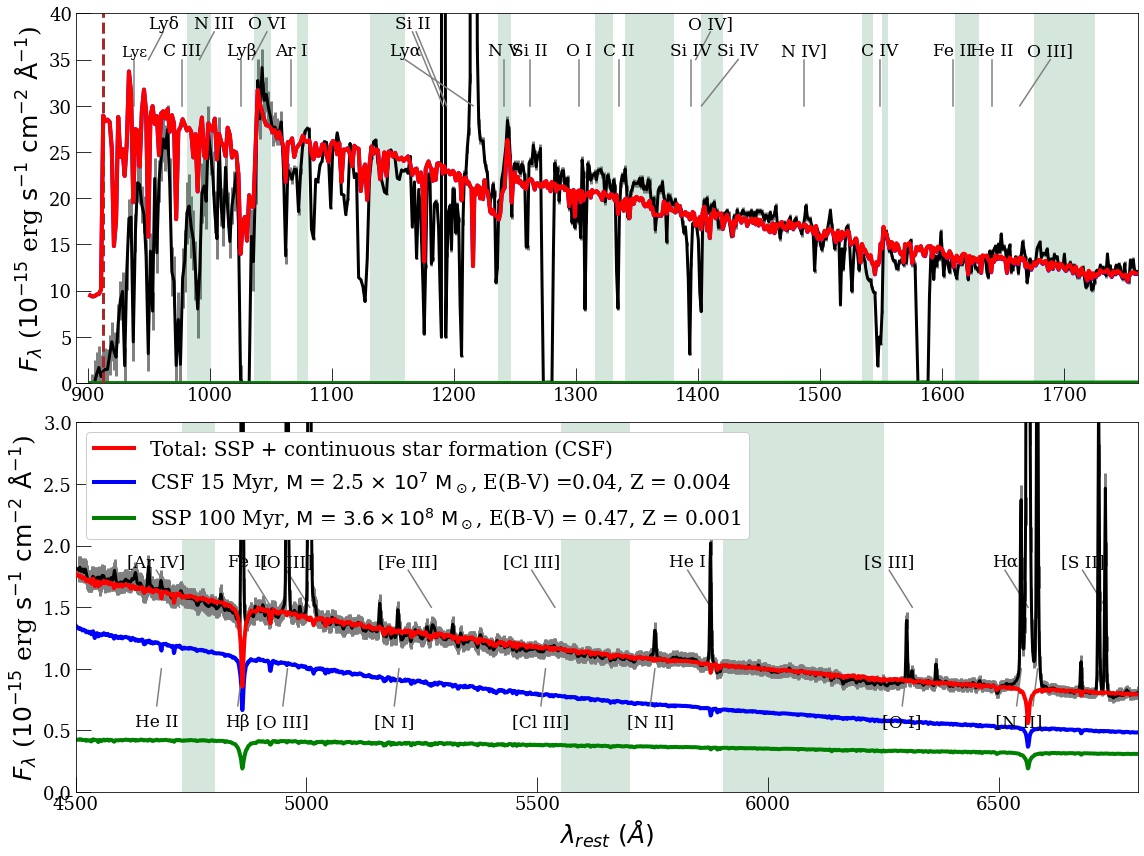}{0.9\textwidth}{}}
\vspace{-15pt}
\gridline{
    \fig{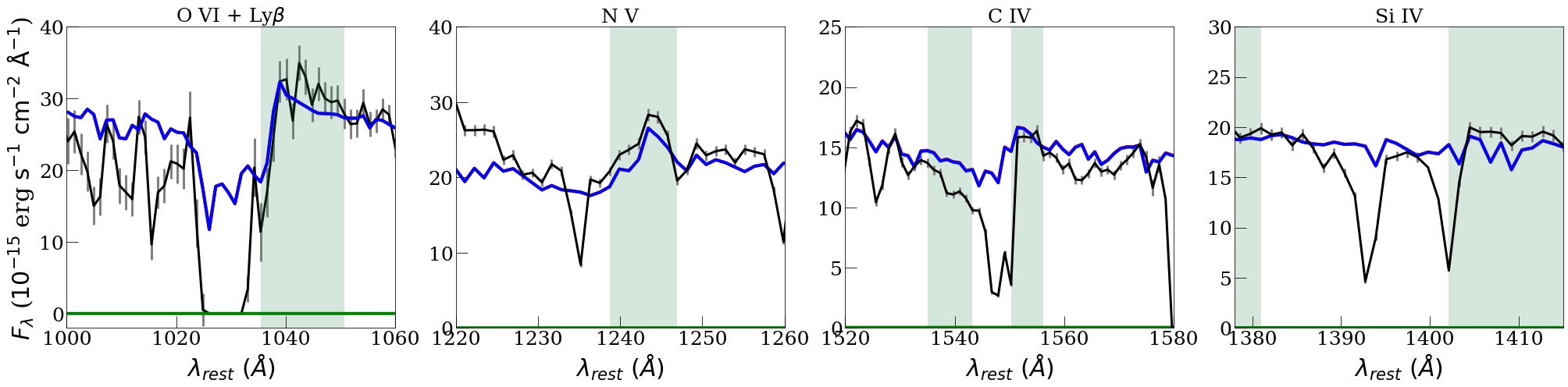}{0.9\textwidth}{}}
\caption{Same as Figure \ref{fig:Astellarpops} but for Knot C. The model assumes one episode of continuous star formation plus an older background population (see text).}
\label{fig:Cstellarpops}
\end{figure*}

\begin{deluxetable*}{lcccc}
\tablecolumns{5}
\tablecaption{LyC Fluxes and Observationally Derived Properties of Knots A, B, and C.
\label{Table2}}
\tablehead{
\colhead{Parameter} & \colhead{Knot A} & \colhead{Knot B} & \colhead{Knot C} & \colhead{Reference}}
\startdata
$F_{912}$ ($10^{-15}$ erg $\rm s^{-1} cm^{-2}$~\AA$^{-1}$) & $< 3.35$    & 2.31 $\pm$ 1.8    & 1.20 $\pm$  0.94     &  1 \\
$L_{\rm 912, obs}$ ($10^{40}$ erg $\rm s^{-1}$) & $< 2.82$  &  1.95 $\pm$ 1.5 & 0.92 $\pm$ 0.72 &  1 \\
12 + log(O/H)  & 8.42 $\pm$ 0.02 &     8.64 $\pm$ 0.02   & 7.64 $\pm$ 0.01 & 2    \\ 
$\rm O_{32}$   & 9      & 8   & 1     & 7 \\
$Q_0(\rm H\alpha)\tablenotemark{\rm \scriptsize a}~(10^{53}~\rm s^{-1})$ & $2.3\pm0.1$  & $8.2\pm0.4$& $2.3\pm0.1$ & 4\\
$f_{\rm c,Si~II}$ & 0.95 $\pm$ 0.05 & 0.96 $\pm$ 0.04 & 0.55 $\pm$ 0.05 & 6 \\
$f_{\rm c,HI}$ & $\sim 1$ & $\sim 1$ & $\sim 0.8$ & 6\\
$\log(N_{\rm Si~II}/\rm cm^{-2})$ & $15.2\pm0.3\tablenotemark{\rm \scriptsize b}$ & $15.3\pm0.3$ & $14.7\pm0.3$ & 6 \\
$\log(N_{\rm HI}/\rm cm^{-2})$ &$20.7\pm0.1\tablenotemark{\rm \scriptsize b}$ & $21.0\pm0.1$ & -- & 6 \\
$L_{\rm Ly\alpha}$ ($10^{40}$ erg $\rm s^{-1}$) & 3.2 $\pm$ 0.3 & 6.1 $\pm$ 0.6 & 14.0 $\pm$ 1.4 & 6 \\
$f_{\rm esc, Ly\alpha}$  & 1.2 $\pm$ 0.1\% & 0.65 $\pm$ 0.06\%  & 5.9 $\pm$ 0.6\%    & 6         \\
$v_{\rm sep, Ly\alpha}$ (km $\rm s^{-1}$) & 532 $\pm$ 20 & 409 $\pm$ 20 & 400 $\pm$ 20 & 6 \\
$L_{\rm X}$ [0.3-8 keV] ($10^{40}$ erg $\rm s^{-1}$)    & --    & 9.5 $\pm$ 1  & 4.5 $\pm$ 1 & 5 \\
\enddata
\tablenotetext{\rm a}{The emission rate of H-ionizing photons, derived from the de-reddened H$\alpha$ luminosity.}
\tablenotetext{\rm b}{We estimate the errors on the published column densities from other measurements in the study, but the uncertainties are poorly constrained due to asymmetric confidence intervals.}
\tablerefs{(1) This work; (2) \citet{Menacho2021}; (3) \citet{James2013}; (4) \citet{Sirressi2022}; (5) \citet{Gross2021}; (6) \citet{Ostlin2021}; (7) \citet{Keenan2017}; (8) \citet{Menacho2019}}
\end{deluxetable*}

\section{Stellar Populations and LyC escape}
\label{sec:Results}

To quantify the LyC escape efficiency in each region, we compute the local escape fraction as:
\begin{equation}
f_{\rm {esc, 912}} = \frac{L_{\rm 912, obs}}{L_{\rm 912, int}} \quad ,
\label{eq:fesc}
\end{equation}
where $L_{\rm 912, obs}$ is the observed integrated luminosity in our 9~\AA~window of 903 -- 912~\AA~at rest (or 922 -- 931~\AA~in the Haro 11 frame), and $L_{\rm 912, int}$ is the intrinsic luminosity in this range, similar to \cite{Leitet}. Thus, \fesc\ is an approximation for the fraction of the total produced ionizing power that escapes the region. We assume that the ionizing luminosity is dominated by FUV radiation from massive stars, and we estimate $L_{\rm 912, int}$ by modeling the stellar population in each knot as described below.

\cite{Sirressi2022} have previously constrained the stellar populations of Haro 11 using multi-band HST photometry in the F140LP, F220W, F275W, F336W, F435W, F550M, F555W, F665N, and F814W filters \citep{Ostlin2009, Adamo2010}. Fitting the resulting SEDs, they estimate individual cluster parameters in each knot. They additionally use FUV $1150 - 1800$~\AA~COS G130M/1300 + G160M/1600 spectra \citep[programs 15352, 13017; PIs Östlin, Heckman]{Ostlin2021} and optical $4650 - 7000$~\AA~MUSE spectra \citep[program 096.B-0923(A); PI Östlin]{Menacho2019}. 
The G130M/1300 spectra were corrected for broad Ly$\alpha$ absorption as described in \cite{Sirressi2022}.
As detailed in \cite{Menacho2019}, the MUSE observations of each knot are the result of 16 dithered integrations at 4 different position
angles. The final MUSE spectra were extracted with apertures of the same size as for the COS spectra,  2.5\arcsec~in diameter, and corrected for vignetting,  with the resulting FUV and optical continua well matched across the wavelength gap.

We combine these existing spectroscopic observations with our newly obtained COS G130M/1055 spectra ($900 - 1200$~\AA) to model each knot's stellar content. 
To increase the signal-to-noise in the UV, we further bin all COS spectra by a factor of 2 in wavelength and median-combine the overlap region. Since the G130M/1055 spectra have different initial extracted spectral sampling from the  G130M/1300 and G160M/1600 data (0.6 and 0.4 \AA\ $\rm px^{-1}$, respectively), the final combined UV spectrum has sampling of 1.2 and 0.8~\AA~$\rm px^{-1}$, over the respective ranges, while the optical MUSE data sampling is 1.2~\AA~$\rm px^{-1}$. The combined spectra of the three knots are shown in Figures \ref{fig:Astellarpops} -- \ref{fig:Cstellarpops}. Below we describe our stellar population modeling assumptions, spectral features of interest, and fitting procedure. 

We use Starburst99 \citep{Leitherer1999, Leitherer2014} model spectra of varying cluster ages ($1 - 100$ Myr), masses ($10^5 - 10^9~\rm M_\odot$), and extinctions ($E(B-V) = 0$ to 1). We also consider different star formation histories, specifically, constant star formation (CSF) or single stellar populations (SSPs). We fit either CSF or a maximum of 3 SSPs to each knot's combined COS G130M/1055 + G130M/1300 + G160M/1600 + MUSE spectrum. While the S22 photometric analysis revealed at least 7 individual star clusters in each knot, our goal is to 
approximate the $1-3$ dominant stellar populations within each region.
We evaluate a number of stellar model assumptions by using a variety of evolutionary tracks. With respect to S22, we extend the parameter space to include stellar rotation, testing Geneva 2012 tracks with $v = 0$, $v = 0.4 \times v_{\rm breakup}$ \citep{E12}. We also consider standard or high mass loss: \cite{E12} or \cite{Meynet1994} tracks. For all models, a Salpeter initial mass function (IMF) with stellar mass range $0.1 - 120~\rm M_{\odot}$ is assumed, similarly to S22. The stellar mass normalizations depend on the low-mass end of the IMF, and we estimate that our resulting masses would be $1.6\times$ lower if a Kroupa IMF had been assumed. We re-sample the model spectra onto our observed wavelength grid, matching the variable sampling, and convolve with a Gaussian of width $\sim 1$ \AA, equal to the stellar broadening we measure from photospheric C III 1247.

Following S22, we perform a linear interpolation of the model spectra between the discrete metallicity values available in Starburst99 ($Z = 0.001, 0.008, 0.02$), in order to estimate models for metallicities appropriate to Haro 11. Previous Haro 11 metallicity measurements have yielded discrepant results spanning a factor of two in each knot \citep{Guseva2012, James2013, Menacho2021}. 
\cite{Menacho2021} provide a comprehensive summary and discussion of the previous metallicity measurements. 
While these measurements relied on the same, direct method, the discrepancies are attributed to differences in ionization correction factors and aperture sizes, as well as some possible real variations between stellar and ionized gas metallicities probed by different apertures. We therefore use the most spatially resolved metallicity measurements for Haro 11 by \cite{Menacho2021}, adopting their measured central values for Knots A and B. For Knot C, which shows a much larger age spread, $1-100$ Myr, we treat metallicity as a free parameter and obtain values consistent with \cite{Menacho2021}. Lastly, we apply extinction to the models, using the SMC law from \cite{Prevot1984}. If the \cite{Calzetti2000} law is assumed, it again results in unreasonably high extinctions compared to previous measurements, and thus also stellar masses.

Since our objective is to constrain the youngest, LyC-emitting population, we focus on fitting the age-sensitive O VI~$\lambda$1038, N V~$\lambda$1243, C IV~$\lambda$1548, and Si IV~$\lambda$1400 P Cygni features, which trace stellar winds from young stars, as well as the optical and UV continuum, which helps constrain the extinctions and mass normalizations. While we do not weigh the P Cygni profiles differently in fitting the spectrum, we reject models that fit the continuum but not the P Cygni features. We mask out the narrow interstellar absorption components before fitting, taking special care in the absorption regions of P Cygni lines. We further isolate feature-free sections of the UV and optical continuum, and mask out nebular emission lines, ISM features, and detector gaps before fitting. The spectral regions we use for fitting are shown in the green sections in Figures \ref{fig:Astellarpops} -- \ref{fig:Cstellarpops}. 
We note that O VI~$\lambda$1038 has contamination from Ly$\beta$ and moreover, the stellar wind feature in the model atmospheres is 
especially uncertain due to combined effects of wind inhomogeneities and X-rays, as well as low signal-to-noise. 
For each knot's spectrum, discrete age combinations from 1 to 100 Myr are fit in steps of 1 Myr, determining the best-fitting masses and extinctions using the differential evolution minimizer in the Python package {\tt lmfit}. The best model is then chosen as the age, stellar mass $M$, and extinction $E(B-V)$ combination that results in the lowest $\chi^2$.\par
Testing the effects of stellar rotation and assumed mass loss rates, we find that, for all three knots, the evolutionary tracks that produce the highest-quality fits are high mass-loss, non-rotating 1994 Geneva models \citep{Meynet1994}. In these tracks, the mass-loss rate was doubled from the ``standard'' case \citep{Schaerer1993}, to match observations. Since 2012, a lower, theoretical mass-loss prescription \citep{Vink2001} has been adopted by the Geneva group accounting for wind inhomogeneities, based on clumping-corrected mass-loss rates \citep{Ekstrom2012, Georgy2013}. Models assuming the opposite cases of fast rotation and low mass-loss produce reasonable fits for Haro 11, but with consistently higher $\chi^2$ values. 
The preference for non-rotating tracks in Haro 11 suggests the rotation rates to be at the slower end of the spectrum defined by the two available rates of $v = 0$ and $v = 0.4 \times v_{\rm breakup}$. Noting that mass loss and rotation rates require further study in low-metallicity systems, we present the stellar population fits assuming high mass-loss and zero rotation, similarly to S22.

For each knot's stellar population fit, we compute $L_{\rm 912, int}$ from the model spectrum and obtain the escape fraction from equation~\ref{eq:fesc}. For the escape fraction uncertainty, we combine the error in the detected LyC luminosity with that in the modelled $L_{\rm 912, int}$, which we obtain by Monte Carlo sampling our best-fit stellar masses and ages $10^4$ times. We also calculate the predicted $Q_{0}$, the emission rate of H-ionizing photons, to compare to that inferred from H$\alpha$ observations, $Q_0(\rm H\alpha)$. The latter was measured for all knots
by \cite{Sirressi2022} from the MUSE spectra,
extracted with the same apertures as in the FUV COS observations. HST/WFC3 F665N imaging suggests that
$Q_0(\rm H\alpha)$ is underestimated by
$25\%$, $15\%$, and $15\%$ for Knots A, B, and C, respectively. We also estimate the predicted thermal FIR luminosity of each knot that arises from dust-processed optical and UV radiation, by integrating the difference between the intrinsic and the reddened model spectra in our fitting ranges $912-1750$~\AA~and $4500-7000$~\AA. To compare to observations, we derive the $1-1000~\mu \rm m$ luminosity from the galaxy-integrated \textit{IRAS} $F_{60 \mu \rm m}$, $F_{100 \mu \rm m}$ fluxes and the \cite{Helou1988} prescription, obtaining $L_{\rm FIR} = 2.7\pm1.3\times10^{44}~\rm erg~s^{-1}$. So, the dust content in Haro 11 generates a substantial FIR luminosity consistent with luminous infrared galaxies (LIRGs). 
This value is reasonably consistent with the sum of our
estimated FIR luminosities from the individual knots obtained below, which is $1.7\times10^{44}~\rm erg~s^{-1}$.

Detailed stellar population fits for each knot are presented below, with resulting parameters shown in Table~\ref{Table3}, together with a discussion of the LyC escape efficiency and mechanisms. We present our results for the three knots in order of decreasing detected LyC flux. 

\begin{deluxetable*}{lccccccccccccc}
\tablecolumns{14}
\tablewidth{0pt} 
\tablecaption{Parameters Fitted from Spectroscopic Stellar Population Models}
\label{Table3}
\tablehead{\colhead{Region} & \colhead{Model\tablenotemark{\rm \scriptsize a}} & \colhead{Age} & \colhead{$M$} & \colhead{$E(B-V)$} & \colhead{$Z$\tablenotemark{\rm \scriptsize b}} & \colhead{$L_{\rm FIR}$} & \colhead{$Q_{0}$} & \colhead{$\log(\xi_{\rm ion})$}& \colhead{$f_{\rm esc, 912}$} \\
\colhead{} & \colhead{} & \colhead{Myr}& \colhead{$10^6~\rm M_\odot$} & \colhead{} & \colhead{} & \colhead{$10^{43}~\rm erg~s^{-1}$} & \colhead{$10^{53}~\rm s^{-1}$}&  \colhead{$\rm Hz~erg^{-1}$}& \colhead{}}
\startdata
{} &{} & 3$\pm0.5$ &7.4$\pm0.6$ &  0.05$\pm0.01$ & {} &0.8 &2.3 &{} &{} \\
\rm Knot~A & \rm 3 SSPs & 4$\pm0.5$ & 13.4$\pm4.1$ & 0.46$\pm0.2$ &  0.011 & 1.9 &2.2 & 25.23 & $< 0.10$ \\
{} & {}& 13$\pm3$ & 4.5$\pm1.2$ & 0.16$\pm0.05$ & {} &0.1& $3\times10^{-3}$ & {} &{} \\
\hline
{} & \rm CSF & 3$\pm0.3$ & 1.9$\pm 0.4$ & 0.04$\pm 0.02$ & {} & 0.3 & 0.7 & {} & {}  \\
\rm Knot~B & \rm SSP & 1$\pm0.2$ & 22$\pm1.5$ & 0.5$\pm0.1$ &  0.018 & 7.7 & 9.3 &  25.32 & 0.034$\pm0.029$  \\
{} &SSP & 13$\pm2$ & 66$\pm10$ & 0.21$\pm 0.06$ & {} &4.7 & 0.02 & {} &  {} \\
\hline
\rm Knot~C &\rm CSF & 15$\pm3$ &26$\pm4$ & 0.04$\pm0.01$ & 0.004 & 1.1& 4.0& 25.22 & 0.051$\pm0.043$ & \\
{} & SSP & $>30$\tablenotemark{\rm \scriptsize c} & 360$\pm80$ & 0.47$\pm0.25$ & 0.001 & 0.7&$2\times10^{-4}$ &{} & {} \\
\enddata  
~\\
\textbf{Notes.} 
\tablenotetext{\rm a}{Model for star formation: single stellar populations (SSPs) or continuous star formation (CSF).}
\tablenotetext{\rm b}{From \cite{Menacho2021}.}
\tablenotetext{\rm c}{Fits based on both Starburst99 and Yggdrasil (see text). Ages up to $\sim$ 2 Gyr can be fit.}
\end{deluxetable*}

\begin{deluxetable}{ccccc}
\tablecolumns{5}
\tablewidth{0pt} 
\tablecaption{S22 Photometric Clusters
\label{table:clusters}}
\tablehead{
\colhead{Region} & \colhead{Age} & \colhead{$M$} & \colhead{$E(B-V)$} & \colhead{$N_{\rm cl}$\tablenotemark{\rm \scriptsize a}}\\
\colhead{} & \colhead{Myr}& \colhead{$10^6~\rm M_\odot$} & \colhead{} & \colhead{}}
\startdata
{} & 4 & 1.4 & 0.06 &1\\
\rm Knot~A  & 5 & 1.7 & 0.11 & 3 \\
{} & 6 & 3.3 & 0.06 & 3 \\
{} & 14 & 4.7 & 0.02 & 1 \\
{} & 15 & 10.8 & 0.07 & 3 \\
\hline
{} & 1 & 8.7 & 0.5 & 2\\
\rm Knot~B &  2 & 7.2 & 0.4 & 1\\
{} & 4 & 13.7 & 0.44 & 2 \\
{} & 5 & 44.1 & 0.4 & 4 \\
{} & 14 & 1.6 & 0.3 & 1 \\
\hline
\rm Knot~C & 6, $8-10$ & 2.2 & 0.06 & 4\\
{}& 15 & 44.3 & 0.05 & 3
\enddata  
\textbf{Notes.} 
\tablenotetext{}{S22 cluster parameters combined in age bins, where masses are summed for the bin, and extinctions are averaged. The typical uncertainties on ages, masses, and extinctions are 1 Myr, $0.5 \times 10^6~\rm M_{\odot}$, and 0.02, respectively. The metallicity fixed for the cluster analysis is $Z = 0.004$.}
\tablenotetext{\rm a}{$N_{\rm cl}$ is the number of clusters in each age bin.}
\end{deluxetable}

\subsection{Knot B}
\label{sec:KnotBStellarpop}

We find that the region of Haro 11 with the brightest measured LyC is Knot B. Using the distance of 88.5 Mpc to Haro 11, we obtain $L_{\rm 912,obs}$ = $(2~\pm~1.5) \times 10^{40}~\rm erg~s^{-1}$ for Knot B (Table~\ref{Table2}). 
This region by far dominates the H$\alpha$ emission of Haro 11 and thus has the highest intrinsic ionizing power among the three knots, with $Q_0(\rm H\alpha) = 8\times10^{53}~\rm s^{-1}$ \citep{Sirressi2022}. Its young massive stars should therefore dominate the production of ionizing photons in Haro 11. However, this object shows large amounts of gas and dust, and it therefore has been somewhat overlooked in the literature when considering the origin of the LyC from this galaxy.

Our stellar population synthesis confirms that Knot~B produces the most ionizing radiation in Haro 11. The best-fitting model for the combined COS G130M/1055 + G130M/1300 + G160M/1600 + MUSE spectrum of Knot B is shown in Figure \ref{fig:Bstellarpops}. A strong match for the spectrum (reduced $\chi^2  = 1.7$), with age-sensitive P-Cygni wind features well fitted,
consists of a continuous star formation episode for the last 3 Myr, as well as 1-Myr and 13-Myr single stellar populations. The 1~Myr component is the youngest age observed in Haro~11, as can be seen in Table \ref{Table3}, where detailed stellar population model parameters are shown for all knots.
Moreover, Knot B has the most massive LyC-bright population, where the stellar mass in ages $<5$~Myr is $2.4\times 10^7$~\msun, which is $3\times$ higher than that found in the second-brightest LyC-leaking region, Knot C. As shown in Table \ref{Table3}, Knot B thus has the highest ionizing photon production efficiency among the knots, $\log(\xi_{\rm ion}) = 25.3$, compared to 25.2 in Knots A and C. Notably, the dominant, 1-Myr component in Knot B is heavily obscured, showing $E(B-V) = 0.5$ and contributing marginally to the observed FUV flux. 
The age-sensitive P Cygni features are thus accounted for by the 3 Myr CSF component alone, while the dusty 1 Myr component contributes $\sim30\%$ of the optical luminosity. But with a mass of $\sim2\times10^7~\rm M_{\odot}$, the 1 Myr population successfully reproduces the high photoionization rate inferred from H$\alpha$. As can be seen in Tables \ref{Table2} and \ref{Table3}, our model gives a total $Q_{0} = 1\times10^{54}~\rm s^{-1}$, while that inferred from H$\alpha$ is $Q_0(\rm H\alpha)$ = $0.8\times10^{54}~\rm s^{-1}$, which is in good agreement. Our predicted FIR luminosity, dominated by the 1-Myr component and tracing dust-processed radiation, is $L_{\rm FIR} = 1.3\times10^{44}~\rm erg s^{-1}$. Our modeling thus confirms the existence of a massive, obscured young population that dominates the ionizing power and FIR luminosity of Knot B.

Our results can be compared to the S22 photometrically derived cluster parameters given in Table \ref{table:clusters}.
S22 photometric modelling pointed to 10 clusters within the COS aperture with ages 1--5 Myr of varying masses and extinctions, as well as a 14 Myr cluster. Since we fix the number of separate population components in our spectroscopic analysis to three, our model for Knot B cannot account for all the S22 cluster ages. But our model is consistent with the two youngest S22 clusters, 1~Myr and 2~Myr, where our total mass in these ages is within 40\% of the corresponding mass found by S22, and we find similarly high extinctions of $E(B-V)\sim 0.5$. 
Notably, while Wolf-Rayet (WR) emission features (e.g., He~I~$\lambda$1640, and the $\lambda$5608~\AA\ and $\lambda$4650~\AA~bumps) are not included in Starburst99 models, these features observed in the spectrum of Knot B are signatures of a $\gtrsim$3~Myr population, consistent with the presence of photometrically derived 4 Myr old clusters (Table \ref{table:clusters}). Our 3-Myr CSF component thus accounts both for the presence of the youngest stars, as well as the more evolved WR components. Lastly, the 13 Myr component we find is $40\times$ more massive than the 14 Myr cluster, which may be attributed to the spectroscopic aperture also capturing the extended, diffuse populations. 

Given the large cluster masses and young ages
seen in Knot B, it could feasibly host Very Massive Stars (VMS, $\rm M > 120~M_\odot$). We examine the spectrum of Knot B for VMS diagnostics, such as O V $\lambda$1371 absorption, He II $\lambda$1640 emission with equivalent width (EW) $> 3~$\AA , an absent or weak double-peaked red bump, and a blue bump without  WR lines \citep{Kunth1981, Martins2023, Wofford2023}. We do not detect O V $\lambda$1371 in Knot B, and we measure its He II $\lambda$1640 EW to be $2.3$~\AA. The red bump detected in Knot B is broad and smooth, while the blue bump shows WR C IV $\lambda$4658. The above are all consistent with classical WR stars, and not VMS.

Although Knot B is the brightest LyC source in Haro 11, its local escape fraction is low. Based on the intrinsic LyC luminosity of our modelled stellar components, we obtain \fesc = $0.034$$\pm 0.029$ (Table \ref{Table3}). The uncertainty is dominated by observational error of $77\%$ on the detected flux, while the uncertainty in the intrinsic LyC luminosity is $21\%$. To compare to the escape fraction implied by the component cluster parameters, we model the clusters' UV spectra in Starburst99, based on their ages, masses, and metallicities. Summing their intrinsic LyC luminosities in the $903-912$~\AA~range, we obtain \fesc = $0.02 \pm 0.01$ for the clusters, consistent with our spectroscopic model.  Knot~B also shows the lowest \lya~escape fraction among the three knots, $f_{\rm esc, Ly\alpha} =$ 0.65 \% \citep{Ostlin2021}, although the observed \lya~peak separation of 400 $\kms$ is similar to that observed in Knot C and other weak LyC leakers with $f_{\rm esc,LyC}\sim0.03$ \citep{Izotov2018, Flury2022b}. The inefficiency of LyC and \lya~escape in Knot B can be explained by its significant gas and dust reservoir. It exhibits the highest H~I column densities among the three knots, $\log(\rm N_{HI}/cm^{-2}) \approx 21$ \citep{Ostlin2021}, and the highest molecular gas mass, $M_{\rm H_2}$ = 2 x $10^9$ $\rm M_\odot$ \citep{Gao2022}. Moreover, it has a neutral covering fraction in Si II, $f_{\rm c,Si~II} = 0.96$ \citep{Ostlin2021} and thus $f_{\rm c,HI} \geq 0.96$ in H~I \citep{Chisholm2018}, consistent with the escape fraction we obtain. With filamentary dust clouds across the region further obscuring it, Knot B shows significant dust extinction of $E(B-V) = 0.5$ (Table \ref{Table3}).  

Ionization-parameter mapping (IPM), on the other hand, points to a complex picture of optical depth. Knot B shows a significant [S II]~$\lambda\lambda$6717,6731 deficiency, $\Delta[\rm S~II] =  -0.16$ (\"{O}stlin et al., in prep), where $\Delta[\rm S~II]$ is the displacement in $\log([\rm S~II]/H\alpha)$ from typical star-forming
galaxies
\citep{Wang2021}, which may indicate density-bounded conditions \citep{Pellegrini2012,Wang2021}. Values of [O II]~$\lambda$3727/H$\alpha < 0.1$ indicate low optical depth, and for Knot B, [O II]~$\lambda$3727/H$\alpha \sim 0.05$ in the central line of sight
\citep{Keenan2017}, suggests LyC escape. Knot B also shows elevated $\rm O_{32}$ = [O III]~$\lambda$5007/[O II]~$\lambda 3727 >8$ overall \citep{Keenan2017}, roughly twice the mean value observed in local unresolved LCEs \citep{Flury2022a}. Higher $\rm O_{32}$ values point to density-bounded conditions and correlate with the LyC escape fraction \citep{Izotov2016a, Izotov2016b, Flury2022b}. However, the extended region shows a confined, ionization-bounded morphology in transverse directions \citep{Keenan2017}. Therefore, Knot B must be leaking LyC through a narrow ionization cone in the line of sight. \cite{Menacho2019} report evidence of narrow highly ionized channels with Knot B at the base. They also find 1000 km/s outflows, likely driven by stellar feedback. We measure $\rm [S~II]\lambda\lambda6716,6731/H\alpha = 0.14$ in our spectrum of Knot B, too low to signal shock heating by supernovae. The S22 photometrically derived cluster parameters in Knot B, the youngest of which we capture in our spectroscopic model, nevertheless imply a total power of $2\times10^{41}$ erg/s in stellar winds and supernovae.

Since the LyC-leaking Knot B hosts an ultra-luminous X-ray source (ULX), its potential contribution to the LyC emission needs to be evaluated. This is an unusually bright, hard ULX \citep{Prestwich2015}. 
The X-ray emission has been revealed to originate from at least two objects \citep{Gross2021}. Based on the X-ray hardness, \cite{Gross2021} suggest that one or both of the sources is a black hole binary in a low-accretion, hard state, with the high X-ray luminosity suggesting the presence of an intermediate-mass black hole (IMBH) of mass $\rm M_\bullet > 7600~M_\odot$ in the region. Alternatively, it may be a low-luminosity AGN (LLAGN), whose signatures are obscured by the dense gas and dust observed in Knot B or diluted by the intense star formation. It is thus possible for Knot B to be a unique merger site of two IMBH's or LLAGN \citep{Gross2021}. 

However, we find that the ULX is unlikely to be sufficiently bright to contribute to the LyC leakage from Knot B. A generous upper limit to the LyC luminosity of the ULX can be estimated by extrapolating the X-ray power-law into the UV. Using the observed slope of $\Gamma = 1.7$ \citep{Gross2021}, we estimate the intrinsic $903-912$~\AA~luminosity of the ULX to be $L_{912,\rm ULX} \lesssim 6\times10^{39}\rm~erg~s^{-1}$. This is fainter than the observed LyC of Knot B, $L_{912,\rm obs} = 2 \times10^{40}\rm~erg~s^{-1}$, although it agrees within the observational uncertainty of $\pm 1.5\times10^{40}\rm~erg~s^{-1}$. Besides, more realistic ULX SED models predict even fainter LyC luminosities than we estimate, by 1-2 orders of magnitude \citep[e.g.,][]{FernandezOntiveros, Gierlinksi2009}.
Regardless, the intrinsic stellar LyC emission we constrain from our population synthesis, $L_{912,\rm int} = 6\times10^{41}\rm~erg~s^{-1}$, is two orders of magnitude higher than that estimated for the ULX. So, the stellar population alone can fully account for the observed LyC leakage from Knot B, with a relatively insignificant contribution from the ULX. 

\subsection{Knot C}
\label{sec:KnotC}
Knot C is the region with the second-strongest LyC flux of Haro 11, with a detected LyC luminosity $L_{\rm 912,obs}$ = $(0.9 \pm 0.7) \times 10^{40}~\rm erg~s^{-1}$ (Table~\ref{Table2}), or nominally about half of the Knot B luminosity. We note, however, that the large uncertainties on the LyC fluxes preclude a conclusive claim on the relative strengths of the emergent LyC from the knots. Knot C has been the prime candidate for LyC emission from Haro 11 based on its highest \lya~escape among the three knots, with $f_{\rm esc, Ly\alpha} =$ 6 \% for Knot C, and $f_{\rm esc, Ly\alpha} \lesssim 1 \%$ for A and B \citep{Ostlin2021}. 
Why does LyC appear fainter than that in Knot B, and what is the corresponding local escape fraction? 

Our best stellar population fit to the combined COS G130M/1055 + G130M/1300 + G160M/1600 + MUSE spectrum of Knot C is shown in Figure \ref{fig:Cstellarpops}, with detailed parameters in Table \ref{Table3}. It consists of two components: a population continuously formed for 15 Myr at $\rm SFR = 1.7~M_\odot~yr^{-1}$ and a single older population with age 100 Myr. The model fits the observed spectrum well (reduced $\chi^2  = 2.4$), with the P Cygni profiles of O~VI, N~V, C~IV, and Si~IV fit well, excluding ISM absorption, and clearly indicating the presence of $1-5$ Myr-old stars. As can be seen in Tables \ref{Table2} and \ref{Table3}, our stellar population fit gives a total photoionization rate $Q_{0} = 4\times10^{53}~\rm s^{-1}$, which is within a factor of two of the value inferred from H$\alpha$. As Knot C appears to be a nuclear star cluster, a continuous star formation history is reasonable \citep{Adamo2010}. 

The current episode of constant star formation is superimposed on an older, background  population, which is likely the diffuse, extended, bulge-like component. We allow the metallicities of the components to vary between observed values, $Z_{\rm obs} = 0.004$ from \cite{Menacho2021}, and $Z = 0.001$. We note that the old component we find is a generic background population whose age is not well determined above $\sim30$ Myr. Because its contribution to the spectrum is only in the optical continuum, it can be fitted with a wide range of age--extinction combinations. The high-resolution Starburst99 UV models are only available up to ages of 100 Myr. We have therefore also fit low-resolution Yggdrasil SSP models \citep{Zackrisson2011} to the spectrum of Knot C and found a good fit with ages up to 2~Gyr and extinctions $E(B-V) < 0.8$, consistent with previously reported values \citep{Sirressi2022}. Our predicted FIR luminosity from dust processing is $L_{\rm FIR} = 2\times10^{43}~\rm erg~s^{-1}$.  

Comparing our results to the S22 photometrically derived cluster parameters in Tables \ref{Table3} and \ref{table:clusters}, 
we see that S22 find Knot C to be dominated by one massive cluster of age 15 Myr, which is likely the nuclear cluster.
The cluster properties we obtain from the spectrum are consistent with S22 within the errors. Thus Knot C, undergoing continuous star formation for the last 15 Myr, has the oldest mass-weighted age in Haro 11, in contrast to the 2 Myr-dominated Knot B. While there are $1-5$ Myr stars present, their total mass in our model is $\sim 9\times10^6~\rm M_\odot$, which is $3\times$ lower than that in Knot B. Knot C therefore has a lower intrinsic ionizing luminosity than Knot B. 

Our stellar population model gives a local escape fraction from Knot C of \fesc = $0.051$$\pm 0.043$.
Here, the uncertainty is dominated by observational error of $78\%$, while the model uncertainty is $30\%$. The escape fraction of Knot C may thus be higher than that of Knot B, $0.034 \pm 0.029$ (Table \ref{Table3}). So, despite appearing fainter in LyC than Knot~B, Knot~C may leak LyC more efficiently.
The observed properties of Knot C are also consistent with it having the highest LyC escape among the knots. In addition to the highest \lya~escape fraction, Knot C exhibits the lowest neutral covering fraction $f_{\rm c,Si~II} <0.5$, corresponding to $f_{\rm c,HI} \sim 0.8$ \citep{Chisholm2018} and the lowest neutral column density, as measured with the apparent optical depth method \citep{Savage1991} applied to Si II, $\log(N_{\rm Si~II}/\rm cm^{-2}) = 14.7$ (Table \ref{Table2}) \citep{Ostlin2021} . 

On the other hand, its extinction is comparable to that of Knot B, suggesting a weak relation between dust extinction and LyC escape. Ionization-parameter mapping results are likewise ambiguous. On one hand, Knot C shows a strong [S~II]~$\lambda\lambda$6717,6731 deficiency of $\Delta[\rm S~II] =  -0.14$ (\"{O}stlin et al., in prep), suggestive of LyC escape \citep{Wang2021, Pellegrini2012}. On the other hand, the knot appears to be in a low ionization state, based on low $\rm O_{32}\le 3$ and [O~III]~$\lambda$5007/H$\alpha \le 0.5$ and high [O~II]~$\lambda$3727/H$\alpha \sim 0.3$ \citep{Keenan2017}. As shown in Figure 5 of \cite{Keenan2017}, it appears to have a high-ionization region extending to the east, but overall, Knot~C appears to have low ionization relative to the rest of the galaxy. It moreover shows a confined morphology for the high-ionization region, with the $\rm O_{32}$ ratio transitioning smoothly and quickly to lower values into optically thick envelopes.  

Despite some signatures of high optical depth, Knot C has the advantage of extensive stellar feedback. With a continuous star-formation history of 15 Myr (Table \ref{Table3}), Knot C has had a significant supernova history. S22 suggest that the mechanical feedback has been taking place even longer, over the last 40 Myr \citep{Sirressi2022}. The S22 feedback model can account for the energetics of the observed soft diffuse X-ray emission in Haro 11 reported by \cite{Grimes2007}. We measure $\rm  [S~II]\lambda\lambda6716,6731/H\alpha = 0.26$ in our spectrum of Knot C, consistent with stellar photoionization. But \cite{Menacho2019} find that the combination of $\rm [O~I]\lambda6300/H\alpha$ and $\rm [O~III]\lambda5007/H\alpha$ ratios on the outskirts of Haro 11 indicates $200-600$ km/s shocks. They also find a $\sim 2$ kpc, high-ionization structure with Knot C at the center, that is likely a superbubble.
Knot C also shows 1000 km $\rm s^{-1}$ gas \citep{Menacho2019}, which is difficult to explain with supernovae or stellar winds, and may instead be a signature of radiation-driven outflows \citep{Komarova2021}.
Overall, the significant stellar feedback in Knot C, whether radiation- or mechanically dominated, may have cleared optically thin channels in its ISM, through which LyC photons can escape. 

Similar to Knot B, Knot C contains a ULX that, based on its $0.3-8.0$ keV spectrum and luminosity, might be one of the most luminous soft ULXs known, with $L_{\rm X} = 4.5\times10^{40}~\ergs$  \citep{Prestwich2015, Gross2021}. The X-ray emission likewise originates in two point sources, where the secondary object shows $L_{\rm X} \sim 2\times10^{40}~\ergs$ \citep{Gross2021}. \cite{Prestwich2015} and \cite{Gross2021} explain its high luminosity by an IMBH of mass $\rm M_\bullet > 20~M_\odot$, undergoing super-Eddington accretion \citep{Swartz2011, Kaaret2017}, although it is also possible to be a neutron star binary, since compact object masses are poorly constrained. In turn, the blowout of inner-disk material in this intense accretion phase can result in a disk wind \citep[e.g.,][]{Middleton2015}. If the super-Eddington accretion drives an outflow with a mechanical luminosity similar to its X-ray luminosity \citep[e.g.,][]{Justham2012}, the outflow power would be comparable to the stellar wind power of Knot C, estimated to be $6\times10^{40}~\ergs$  \citep{Sirressi2022}. So, the ULX feedback may contribute significantly to gas clearing, promoting LyC escape. 

To understand the role of the ULX in the LyC leakage from the knot, we estimate the ULX LyC output by extrapolating its X-ray power-law into the UV. For the observed spectral index $\Gamma = 2.1$, we estimate $L_{912,\rm ULX} \le 2\times10^{40}\rm~erg~s^{-1}$. This is a generous upper limit, as \cite{Vinokurov} and \cite{Kaaret2009} show that more realistic model SEDs of supercritical accretion disks are even fainter in the UV. Our estimated ULX LyC luminosity is twice as bright as the observed LyC luminosity from Knot C, and thus the ULX can plausibly contribute LyC. However, our modelled stellar LyC of $2\times10^{41}~\ergs$ exceeds that of the ULX by at least an order of magnitude. Thus, if the ULX is responsible for some of the LyC emission from the region, its contribution may be unimportant compared to that of the stellar population.  However, if its mechanical feedback dominates the gas clearing for the observed LyC, its LyC contribution may still be significant.

So, the stellar population of Knot C can alone account for the observed LyC leakage from this region. The soft, luminous ULX observed in this knot may be sufficiently bright to contribute to the LyC emission, but its intrinsic production is $<10$\% of  the stellar emission. The ULX may be able to aid in the escape of Lyman radiation through mechanical feedback. Further investigation of the mechanical and radiative feedback of the ULXs is needed to conclusively establish their roles in the LyC leakage from Knot C.

\subsection{Knot A}
\label{sec:KnotA}
Knot A has also been predicted to be the LyC-leaking knot, based on ionization-parameter mapping, which \citet{Keenan2017} use to show that Knot A is responsible for a large, $\sim$kpc-sized region with high O$_{\rm 32}$. They find a central $\rm O_{32} \sim 9$, consistent with the most extreme Green Peas, which are the largest class of local LyC leakers \citep[e.g.,][]{Flury2022a}. However, we do not detect LyC in Knot~A. The $2\sigma$ upper limit of the LyC flux density from Knot A is $F_{\rm 912} < 3.3 \times 10^{-15}~\ergs \rm cm^{-2}$~\AA$^{-1}$ in the range $903-912$~\AA, or $L_{\rm 912, obs} < 2.8 \times 10^{40}~\ergs$. 

To estimate the local escape fraction upper limit, we use our population synthesis results. Our stellar model for the spectrum of Knot A consists of three SSPs of ages 3 Myr, 4 Myr, and 13 Myr, as shown in Figure \ref{fig:Astellarpops} and detailed in Table \ref{Table3}. The model fits the spectrum reasonably well (reduced $\chi^2 = 3.7$), accounting for both the continuum and age-sensitive features. The P Cygni profiles of O VI, N V, C IV, Si IV clearly indicate the presence of 3 Myr-old stars. Although the Wolf-Rayet He~I~$\lambda$1640 and  5608~\AA~and 4650~\AA~bumps are not included in Starburst99 models, these features are consistent with a $\sim$4 Myr population. With $1\times10^7~\rm M_{\odot}$, this 4 Myr component dominates the stellar mass of Knot A, but is highly obscured, with $E(B-V) = 0.5$. It therefore has no FUV contribution, including in the age-sensitive P Cygni features, and the 3 Myr component can alone account for these profiles, while the dusty population contributes $25\%$ of the optical luminosity. The 3 Myr and 4 Myr populations  reproduce the observed photoionization rate from H$\alpha$ within a factor of two (Tables \ref{Table2} and \ref{Table3}). However, given that Knot A has the least dust among the knots, with previously reported extinction $E(B-V)\sim0.2$ \citep{Menacho2021}, this discrepancy may indicate a modest mass overestimate in the 4-Myr component. Comparing with the S22 clusters in Table \ref{table:clusters}, our 4-Myr spectroscopic component is $10\times$ more massive than the 4-Myr S22 cluster, but has an $8\times$ higher extinction. Similar to the 14 Myr cluster found by S22, we find a 13 Myr component, with a matching mass but $2\times$ higher extinction. Lastly, the total FIR luminosity predicted by our model for the knot is $2.8\times10^{43}~\rm erg\ s^{-1}$.

The massive young populations we uncover in Knot~A may possibly host VMS. This was previously suggested as an explanation for Knot A’s broad blue bump \citep{Keenan2017}. Similar to Knot B, we evaluate the spectrum of Knot A for VMS signatures. We do not detect the blue-shifted O V $\lambda$1371 absorption common in VMS, and we measure the He II $\lambda$1640 EW =  2~\AA, while VMS show $>3$~\AA~\citep{Martins2023}. The observed red bump is broad and smooth, while the blue bump shows WR C IV $\lambda$4658. So, as in Knot B, we do not find evidence of VMS in Knot A. The above observations are instead consistent with
classical WR stars.

From our modelled intrinsic LyC luminosity of the stellar populations, we estimate a 2$\sigma$ upper limit to the escape fraction \fesc $\le 0.10$ (Table \ref{Table3}). Our non-detection thus does not conclusively rule out significant LyC escape in Knot A. Indeed, multiple lines of evidence point to some degree of LyC escape in Knot A. First, despite showing the lowest observed Ly$\alpha$ luminosity among the knots, $L_{\rm Ly\alpha} = 3 \times10^{40}~\ergs$, the Ly$\alpha$ escape fraction of Knot A is $f_{\rm esc, Ly\alpha} = 1.2\pm0.12$\% \citep{Ostlin2021}, which is $2\times$ that of Knot B (Table \ref{Table2}). Nevertheless, this value is lower than that suggested by its reddening, implying that \lya~is strongly attenuated by dust scattering \citep{Ostlin2021}. The neutral covering fraction is similar to that of Knot B, $f_c = 0.95\pm0.05$ \citep{Ostlin2021}, and so is the column density, log($\rm N_{HI}/cm^{-2}$) = 20.7 \citep{Ostlin2021}. 

Moreover, Knot A shows a number of signatures of low optical depth in ionization parameter mapping. Its [S~II] deficiency is most extreme among the knots, $\Delta[\rm S~II] =  -0.19$ (\"{O}stlin et al., in prep), predicting the highest $f_{\rm esc, LyC}$ \citep{Wang2021}.  
Most importantly, \cite{Keenan2017} report high-$\rm O_{32}$ gas originating at Knot A and spreading to distances of $>$ 2 kpc, and \cite{Menacho2019} observe it to $>$ 4 kpc as high [O~III]/H$\alpha$, signaling a transparent medium in the transverse directions. In fact, the $\rm O_{32}$ ratio does not transition smoothly into lower values at the edges of this region, strongly implying low optical depth in the plane of the sky. \cite{Menacho2019} show that this structure exhibits the highest ionization, $\rm [O~III]/H\alpha \gtrsim 3$, in velocity bins from $-300~\kms$ to $-150~\kms$, while the central $\rm [O~III]/H\alpha \sim 1$. This suggests that the high-ionization structure is likely an optically thin outflow driven by LyC from the knot, but its axis is not coincident with our line of sight. 

The main implication of our results in light of IPM observations is therefore that LyC escape must be highly anisotropic. While we do not detect LyC in Knot A, it is likely that the leakage, if any, occurs through a channel not coincident with our line of sight \citep[e.g.,][]{Zastrow2011}. This is consistent with the Ly$\alpha$ peak velocity separation, which is highest in Knot A, $v_{\rm sep, Ly\alpha} \sim 500$ km $\rm s^{-1}$, compared to the other two knots.  Such large $v_{\rm sep, Ly\alpha}$  is consistent with escape fractions $<1\%$ \citep{Flury2022b}.

\section{Discussion}
\label{sec:Discussion}
We find that the LyC-emitting regions in Haro 11 are Knots B ($L_{\rm 912,obs}$ = $2.3\pm1.8 \times 10^{40}~\rm erg~s^{-1}$) and C ($L_{\rm 912,obs}$ = $0.9\pm0.7 \times 10^{40}~\rm erg~s^{-1}$). We determine their respective LyC escape fractions to be \fesc$= 3.4\pm2.9\%$ and $5.1\pm4.3$\% (Table~\ref{Table3}). The total LyC-luminosity-weighted escape fraction of Haro 11 is $f_{\rm esc, 912}=3.9\pm3.4\%$, consistent with $3.3\pm0.7$\% obtained by \cite{Leitet}. Knot B appears to dominate in LyC luminosity, with the caveat that the low signal-to-noise in our LyC observations prevents a conclusive determination of which knot dominates. At face value, Knot B is responsible for $2/3$ of the total observed flux, and it has $\sim 3 \times$ higher mass in 1 -- 5 Myr-old stars than Knot C (Table \ref{Table3}), which is the age of peak LyC production. Thus Knot B strongly dominates the ionizing photon production.  However, it has almost full neutral covering \citep{Ostlin2021} and high extinction (Table \ref{Table3}). In comparison, Knot C is a more evolved region of constant star formation for the last 15 Myr, with a commensurate history of supernova feedback, possible LLAGN feedback, and low neutral covering fraction \citep{Sirressi2022, Ostlin2021}. Our findings highlight the sensitivity of LyC escape to the star-formation rate, age, and optical depth.

Thus, Knot B both has more young stars and emits more strongly in LyC, but has a slightly lower escape fraction than Knot C.  Our results underscore the fact that {\it LyC escape fraction and escaping LyC luminosity are separate quantities.}  In Haro~11, Knot~B seems to produce the greatest LyC emission because it strongly dominates in LyC production. But its large gas and dust content apparently keeps its local escape fraction lower than the more evolved, and cleared, Knot C. 
On the other hand, Knot A shows similar age and extinction to Knot B, but we do not detect it in LyC. 
It is thus the interplay of star formation intensity, age, gas clearing, and/or line-of-sight orientation that determines the efficiency and detection of LyC escape. Tracers of \fesc\ alone are insufficient indicators of escaping LyC luminosity along the line of sight.

\cite{Flury2022b} connect LyC properties to star formation density, \lya~properties, and $\rm O_{32}$ in 66 local star-forming galaxies, including Green Peas. They find possible evidence for two modes of LyC escape in the most extreme starbursts, dictated by whether stellar feedback is wind-dominated or radiation-dominated. One population shows younger ages, higher $\rm O_{32}$, and low metallicity, suggesting strong radiation-dominated feedback, which may be linked to optically thin, radiation-driven winds \citep{Komarova2021}. On the other hand, the somewhat older starbursts with lower $\rm O_{32}$ and higher metallicity likely leak LyC with the help of superwinds \citep[e.g.,][]{Heckman2011, Zastrow2013}, and these show lower $f_{\rm esc,LyC}$. 
If Knot A is indeed an anisotropic LyC emitter as evidenced by ionization-parameter mapping \citep{Keenan2017}, then it falls into the radiation-dominated category, consistent with its very high ionization parameter, and other radiation-dominated features \citep[Section~\ref{sec:KnotA};][]{Keenan2017}. Knot B may likewise be radiation-dominated based on its high $\rm O_{32}$ and young age, with the caveat that its metallicity is close to solar. Finally, Knot C has low $\rm O_{32}$ and is apparently dominated by mechanical feedback, given its extensive supernova history \citep{Sirressi2022}. Since we observe multiple stellar generations in each region, it is likely that both modes are at play, with their relative importance to be established. 

Haro 11 is a local Green Pea analog, showing $\rm O_{32}$ and other radiation-dominated properties characteristic of these objects \citep{Micheva2017, Keenan2017}.  These properties are linked to Knot~A, which turns out to not show direct detection in LyC, although IPM strongly implies that the region is optically thin in other sightlines. This additionally stresses the dependence of the escape fraction on the geometrical distribution of dust and neutral gas along the line of sight. The fact that the detected LyC from Haro~11 originates from regions other than Knot~A further demonstrates that LyC emission from Green Peas may be more complex than consideration of a single starburst  \citep[e.g.,][]{Micheva2018}. Notably, the global $\rm O_{32}$ for Haro 11 is only 2.5 \citep{James2013}, which is on the low side for a GP. 
This $\rm O_{32}$ is consistent with that of unresolved GPs in the Low-redshift Lyman Continuum Survey (LzLCS), containing the largest sample of local LCEs to date. The LzLCS GPs show $f_{\rm esc,LyC} = 1-4\%$ for similar $\rm O_{32}$. In addition, the ionizing photon production efficiencies we estimate are $\log(\xi_{\rm ion}) = 25.32$ for Knot B and 25.22 for Knots A and C. These values are lower than that of reionization-era galaxies, $25.4-25.8$ \citep{Simmonds2023, Saxena2023, Atek2024}, or local strong LyC leakers, $25.6-26$ \citep{Schaerer2016}. But they are consistent with the standard values assumed in reionization models \citep{Robertson2015}.

Our spatially resolved study thus uncovers how some of the global properties we observe at higher redshifts may arise from multiple star-forming regions of widely differing properties. In particular, the regions dominating the ionization may not be the primary LyC sources in our line of sight, and more evolved regions' contribution should not be discounted. 

\subsection{Implications for \texorpdfstring{$ Ly\alpha$}{} as a Diagnostic of LyC}

As LyC emission is difficult to observe directly, indirect tracers are required to identify LyC leakers. \lya~is expected to correlate with LyC escape, as it is also sensitive to the hydrogen column density. Radiative transfer simulations \citep[e.g.,][]{Verhamme2015} and observations of several LCEs \citep{Verhamme2017} confirm a tight relationship between Ly$\alpha$ and LyC escape. The LzLCS survey shows that Ly$\alpha$ width and peak separation are some of the strongest indirect predictors of LyC escape fractions \citep{Flury2022b}. With their new, larger sample, the authors reproduce the anti-correlation between Ly$\alpha$ peak velocity separation $v_{\rm sep, Ly\alpha}$ and LyC escape fraction, $f_{\rm esc, LyC}$, first established by \cite{Izotov2018}. 

Our findings in Haro 11 are consistent with these \lya~predictions. As shown in Table \ref{Table2}, the \lya~peak separations observed in Knots A, B, and C are 530, 409, and 400 $\rm km~s^{-1}$, respectively \citep{Ostlin2021}, decreasing with increasing $f_{\rm esc, LyC}$. Knots B and C are consistent with the \cite{Flury2022b} $v_{\rm sep, Ly\alpha} - f_{\rm esc, LyC}$ relation, which predicts $f_{\rm esc, LyC}\sim0.03$ for $v_{\rm sep, Ly\alpha} = 400~\rm km~s^{-1}$. The two LCE knots are also consistent with a $f_{\rm esc, Ly\alpha} - f_{\rm esc, LyC}$ correlation, where Knot C has a higher $f_{\rm esc, Ly\alpha}$ and a slightly higher $f_{\rm esc, 912}$. We calculate the luminosity-weighted average \lya~escape fraction for the three knots to be $f_{\rm esc, Ly\alpha}=4.2\pm0.6$\%. We also estimate the global \lya~peak velocity separation by combining the \lya~profiles of the three knots, obtaining $v_{\rm sep, Ly\alpha} = 410\pm70~\kms$. Our luminosity-weighted LyC escape fraction $f_{\rm esc, 912}=3.9\pm3.4$\% is consistent with the averaged $v_{\rm sep, Ly\alpha}$ according to the \cite{Flury2022b} relation. These values account only for \lya~emission from the knots, and not diffuse \lya~observed outside of them \citep{Ostlin2009}.

The efficiency of \lya~escape, or its escape fraction, can provide insight into LyC radiative transfer. Knot C has both the highest \lya~luminosity and $f_{\rm esc, Ly\alpha}$ among the knots, and its  $f_{\rm esc, 912}$ is indeed likely the highest despite the LyC luminosity being only half of that in Knot B. So, {\it while the LyC and \lya~escape fractions correlate, the emerging LyC and \lya~luminosities do not necessarily do so}. 

As for the shape of the \lya~profiles, Knot A and B clearly show broader \lya~red peaks than Knot C, with their respective widths $302~\kms$, $338~\kms$, and $195~\kms$, suggesting higher optical depth and low LyC escape \citep{Ostlin2021}. The \lya~red peak width specifically points to Knot B as the weakest leaker, inconsistent with our results. Since both Knots A and B show neutral covering fractions close to unity \citep[Table \ref{Table2};][]{Ostlin2021}, it appears that the correlation of \lya~red peak width with $f_{\rm esc, LyC}$ is not as strong when considering individual star-forming regions instead of integrated galaxy properties. The \lya~profile of Knot C, on the other hand, is narrow,  but with a higher red peak asymmetry than in Knots A and B \citep{RiveraThorsen2017, Ostlin2021}. \cite{Kakiichi2021} show that higher red peak asymmetry around its center points to LyC escape through a hole-ridden ISM, or a picket fence structure, as the asymmetry is seen to correlate with ISM porosity in their radiation-hydrodynamic simulations. This arises because the asymmetry of the red peak traces the presence of both optically thin and thick channels, while a symmetric red peak indicates isotropic leakage. 

One inconsistency we see in Ly$\alpha$ predictions is in the non-detected Knot A, where its peak velocity separation is highest among the knots, but its Ly$\alpha$ escape fraction is twice that of Knot B. The peak velocity separation thus implies the lowest LyC $f_{\rm esc, LyC}$~among the knots, while the Ly$\alpha$ escape fraction suggests it should be higher than that in Knot B. A likely explanation for this may be that the 
LyC is emerging in directions transverse to our line of sight, as implied by ionization-parameter mapping (Section~\ref{sec:KnotA}), while \lya\ can be enhanced by scattering into our line of sight.
The observed \lya~luminosity of Knot A is notably the lowest among the knots, and the intrinsic LyC we estimate is $3\times$ lower than that of Knot B (Table \ref{Table3}), while exhibiting a similar covering fraction, column density, and thus similarly broad red \lya~peak. Lastly, it is important to note that the \lya~relations described above were established for unresolved galaxies, which likely also consist of multiple star-forming regions with varying properties. In our spatially resolved study, we connect \lya~profiles from smaller, $\sim$1-kpc apertures to individual knot properties, providing a view of separate components that may make up the integrated \lya~observations. 

Although most of our results agree well with \lya~predictions, there are still significant differences in the escape conditions for LyC vs. Ly$\rm \alpha$. The maximum column density at which Ly$\rm \alpha$ can escape is log($N_{\rm HI}$/$\rm cm^{-2}$) $<$ 13, while the threshold for LyC is log($N_{\rm HI}$/$\rm cm^{-2}$) $<$ 17. So, the four orders of magnitude of difference provide a parameter space where gas can be optically thin in LyC but not in Ly$\rm \alpha$. Another major difference in the escape mechanisms is scattering: Ly$\rm \alpha$ scatters strongly, modulating its escape path and additionally promoting dust absorption relative to LyC. \cite{Dijkstra2016} simulate the radiative transfer of Lyman radiation in multiphase ISM to investigate the relationship between LyC and Ly$\rm \alpha$ escape fractions. They find a positive correlation, as expected, but with significant scatter at higher Ly$\rm \alpha$ escape fractions that is driven by gas covering fraction. The corresponding LyC escape fractions in this region are lower than expected from the correlation, decreasing with higher covering fractions. The scatter extends at least two orders of magnitude, consistent with observations \citep{Flury2022b}, showing that the ISM porosity introduces appreciable stochasticity to the relationship between LyC and Ly$\rm \alpha$ radiation. This can also provide a context for the \lya\ vs LyC observations of Knot~A.

Thus, \lya~properties can provide clues to LyC escape conditions, though not without additional independent tracers. Our Haro 11 study shows that \lya\ peak velocity separation is consistent with it tracing $f_{\rm esc,LyC}$, where the knots fall on the observed relation within scatter. But the \lya~luminosity, red peak width, and escape fraction do not correlate directly with LyC escape in regions of varying gas optical depth and covering. Our results underscore the significant distinctions in \lya~and LyC radiative transfer, and that further study into the effects of ISM morphology and anisotropy of LyC escape is needed.

\subsection{IPM and Anisotropy of LyC Escape}
Ionization-parameter mapping relies on nebular emission line ratios with different ionization potentials to find regions of low optical depth, where high-ionization species dominate. This serves as an indirect tracer for LyC escape \citep{Pellegrini2012}.

Our resolved observations of the LyC-emitting regions of Haro~11 are not fully consistent with predictions from IPM, if isotropic escape is assumed. Most importantly, in Knot A, IPM demonstrates LyC escape conditions transverse to the line of sight \citep[][; Section \ref{sec:KnotA}]{Keenan2017}.
Yet we find no LyC detection in the COS aperture, which probes the line of sight. As noted above, this suggests that LyC escape may be extremely anisotropic. This is also implied from our observations of Knot B, which leaks LyC despite being almost fully covered in neutral gas, implying a narrow escape path. Knot C, on the other hand, is seen to be in a low-ionization state, implying high optical depth in LyC. Yet we find it to have the highest LyC escape fraction. Thus the IPM predictions appear to be sensitive to line-of-sight effects.  

Hydrodynamic simulations of starbursts at $z = 4-6$ by \cite{Cen2015} confirm that the LyC escape fraction depends strongly on the viewing angle. They find that only highly ionized, evacuated channels with small solid angles allow significant LyC propagation. Indeed, the observed scatter in LyC escape fractions with respect to correlated starburst properties at $z \sim 3$ and $z < 0.4$ points to a line of sight effect, where partial ISM clearing results in limited transparent paths \citep{Flury2022b, Nestor2011}. Several nearby starbursts with active stellar feedback likewise exhibit narrow ionization cones \citep{Zastrow2011, Zastrow2013}. The line-of-sight bias is thus crucial to account for in reionization studies at all redshifts. Anisotropy of LyC escape, dictated by the gas-clearing mechanisms such as winds and supernovae, as well as the optical depth and ionization structure, need to be accounted for when using IPM as a predictor for LyC escape. 
This may be even more important if accretion-driven feedback is responsible for the necessary gas clearing.
 
Thus, our observations of Haro 11 offer important data on what information IPM does and does not provide on LyC escape.  While it gives insight into the ionization structure in the plane of the sky, additional tracers probing line-of-sight conditions are required to identify objects in which LyC can be directly detected.  

\subsection{X-ray Sources and LyC Escape}
The two LyC-leaking Knots B and C both host ULXs, while the undetected Knot A is purely star-forming. This interesting coincidence raises the question of the role of accretors in LyC escape, which has been a major problem in cosmic reionization \citep[e.g,][]{Volonteri2009, Madau2015}. Although the ULX LyC contribution is likely negligible in Knot B, the ULX in Knot C, which may be an LLAGN, may contribute to the LyC leakage from the region at the $\lesssim 10$\% level. On the other hand, the focused mechanical feedback that can be expected from these X-ray sources might contribute to clearing optically thin channels for LyC escape in both Knots~B and C. Low-accretion, hard X-ray sources as that seen in Knot~B may produce jets of substantial mechanical power \citep[e.g.,][]{Merloni2013, May2018}. Likewise, the soft, super-Eddington sources, such as the one hosted by Knot~C, can also drive radiation-driven disk winds that may be important \citep[e.g.,][]{Middleton2015}.  These feedback mechanisms may significantly enhance ULX contributions to LyC escape.

The question of the role of accretors in cosmic reionization is all the more important in light of the observed excess X-ray emission in early-universe galaxy analogs. For instance, \cite{Kaaret2011} observe an increased X-ray luminosity per star formation rate $L_{\rm X}/\rm SFR$ in blue compact dwarfs (BCDs), and \cite{Brorby2014} find the BCD X-ray luminosity function to have $10\times$ the normalization observed in solar-metallicity galaxies. Also, \cite{Douna2015} find $10 \times$ more high-mass X-ray binaries (HMXBs) per $\rm SFR$ bin in $< 0.2~Z_\odot$ galaxies than in solar-metallicity objects. Extremely metal poor galaxies (XMPG, $Z < 0.05~Z_\odot$) show more ULXs than higher-metallicity galaxies \citep{Prestwich2013}. Moreover, \cite{BasuZych2013} find that a sample of $z < 0.1$ Lyman break analogs (LBAs), including Haro 11, likewise exhibit a higher $L_{\rm X}/\rm SFR$ than solar-metallicity galaxies. Their interpretation is that lower metallicity results in more luminous HMXBs, as weaker stellar winds lead to more massive compact objects. \cite{Brorby2016} quantify this 
$L_{\rm X}-\rm SFR-Z$ relation, where $L_{\rm X}/\rm SFR$ increases with lower $\log(\rm O/H)$. Finally, \cite{Dittenber2020} find that the majority of local \lya~emitters, and thus candidates for LyC escape, may be driven by ULXs, finding a connection between \lya~escape and HMXBs and/or LLAGN. Thus, metal-poor starbursts in the early universe likely formed an overabundance of X-ray binaries, which may have contributed to the process of reionization. 

From our LyC study of Haro 11, we see that LyC-emitting regions may coincide with ULX sites, but the role, if any, of accretors in ionizing radiation production and mechanical feedback remains to be clarified.

\section{Summary and Conclusions}
\label{sec:Conclusions}
Haro 11 is a key object to understanding cosmic reionization as it is the closest, and first, confirmed local LyC emitter. It is dominated by three star-forming Knots A, B, and C of widely varying properties, and since the original, spatially unresolved detection, it has remained unclear which of these three regions is responsible for the observed LyC emission. 
We therefore obtained new HST/COS G130M/1055 observations of each of the knots in the range $900-1200$~\AA, which reveal that Knots B and C are the LyC emitters toward our line of sight. Their respective $903-912$~\AA~luminosities are $1.9\pm1.5 \times 10^{40}~\rm erg~s^{-1}$ and $0.9\pm0.7 \times 10^{40}~\rm erg~s^{-1}$. So, Knot B seems to dominate the leaking LyC luminosity of Haro 11, accounting for 66 $\pm$ 50$\%$ of the detected flux in $903-912$~\AA, and Knot C accounts for 35 $\pm$ 25$\%$. The total Haro 11 LyC luminosity is $2.9\pm 1.0 \times 10^{40}~\rm erg~s^{-1}$.  

We perform stellar population synthesis to constrain the stellar parameters and thus local LyC escape fractions of each knot. For this, we combine our new COS G130M/1055 with \cite{Sirressi2022} COS G130M/1300 + G160M/1600, as well as \cite{Menacho2019} MUSE spectra. Fitting Starburst99 models to each knot's spectrum, we find that Knot B, the brightest LyC emitter, is dominated by a $\sim1$ Myr, $2\times10^7$~\msun~heavily obscured component. Knot C is best fit with a continuous star formation history for $\sim15$ Myr with 
stellar mass of $3\times10^7$~\msun~and low obscuration. Knot A is dominated by a $\sim4$ Myr, $1\times10^7$~\msun~highly obscured population.  Our modeling thus uncovers massive, young, obscured stellar populations in Knot B and Knot A, which dominate the ionizing photon production in their respective regions but have minimal FUV imprint and contribute $<40\%$ in the optical. We do not see conclusive evidence of Very Massive Stars in any of the knots. Instead, we find clear signatures of classical WR stars in Knots A and B.

The primary tracer of the young populations is the large H$\alpha$ luminosity, which we reproduce with our models within a factor of two or better.  Moreover, Haro 11 shows a large FIR luminosity, $L_{\rm FIR} = 2.7\times10^{44}~\rm erg~s^{-1}$, that qualifies it as a LIRG. Our dusty stellar population models for the three knots combined account for $1.7\times10^{44}~\rm erg~s^{-1}$, which is within 60\% of the observed integrated FIR radiation. Thus the model estimates and observed values are in reasonable agreement, especially considering that the IRAS aperture includes the entire galaxy. \cite{Sirressi2022} photometrically detect $7-11$  clusters in each knot, with ages $1-15$ Myr and masses up to $5 \times 10^7$~\msun~(Table \ref{Table3}). Our 3-component spectroscopic models capture the aggregate young, UV-dominant  populations, as well as the $\sim15$ Myr generation that does not contribute significant LyC. The spectrum of Knot C additionally shows an old background population with an age up to $\sim2$ Gyr. 

The corresponding LyC escape fractions in the range $903-912$~\AA~are \fesc~$= 3.4\pm2.9\%$ for Knot B and $5.1\pm4.3$\% for Knot C. In the case of Knot~A, we place a $2\sigma$ upper limit to the escape fraction of \fesc $\le 10$\%. The luminosity-weighted escape fraction for the entirety of Haro 11 is $f_{\rm esc, 912}=3.9\pm3.4$\%, consistent with \cite{Leitet}.

Our results underscore that the LyC escape fraction (\fesc) and escaping LyC luminosity ($L_{\rm 912, obs}$) are distinct fundamental parameters for characterizing LyC escape. Although we find that the majority of Haro 11's LyC flux likely originates from Knot B, the values above demonstrate that its local escape fraction appears to be lower than that of the LyC-fainter Knot C. The reason is that Knot B has by far the highest ionizing photon production; but it exhibits the largest amount of neutral gas among the knots, and in particular, it has a covering fraction close to unity \citep{Gao2022, Ostlin2021}. This results in a potentially lower escape fraction. On the other hand, Knot C is intrinsically fainter in LyC due to an older mass-weighted age, but it is significantly less obscured. Some of its gas has likely been evacuated by feedback, as its HI covering fraction is $f_{\rm c,HI} \sim 0.8$ and its neutral column density is the lowest among the knots \citep[Table~\ref{Table2};][]{Ostlin2021}. Characterizing LyC emission by only \fesc\ would prioritize Knot C over B, whereas Knot B is in fact the more luminous LyC emitter. Thus, although the escape fraction is often emphasized in the literature, the relevant parameter is the convolution of the escape fraction and the intrinsic LyC luminosity, since both drive the observable LyC luminosity and the number of ionizing photons leaked into the IGM.

We also use our new observations to test \lya~as a LyC escape tracer by comparing our results to \lya-based predictions. We see a correlation of \fesc~with the observed \lya~luminosity, and an inverse relation with \lya~peak velocity separation $v_{\rm sep, Ly\alpha}$, as expected. But the \lya~escape fraction $f_{\rm esc, Ly\alpha}$ does not consistently trace LyC escape. While Knot C has both the highest \lya~and LyC escape fractions, the lowest $f_{\rm esc, Ly\alpha}$ in Knot B incorrectly predicts it to be the weakest LyC leaker. The \lya~red peak width similarly points to Knot C as the strongest leaker and Knot B as the weakest, with the latter not consistent with our observations.  So, the observed \lya~luminosity and peak velocity separation appear to more consistently correlate with LyC escape, while $f_{\rm esc,Ly\alpha}$ and \lya~red peak width are less consistent tracers. However, we stress that these results are based on only three star-forming knots in this galaxy.

There are important implications from the fact that  Knot A is not detected in the LyC, despite it driving the Green Pea properties of Haro 11. Green Peas are the largest class of local LyC emitters, and their radiation properties strongly correlate with $f_{\rm esc, LyC}$ \citep{Flury2022a, Flury2022b}. Knot A has therefore been predicted to be a leaking knot based on its ionization parameter, which is the highest of the three knots \citep{Keenan2017}. First, its non-detection highlights the potential importance of multiple star-forming regions driving the LyC escape in Green Peas. In Haro 11, we see that while the knot dominating the GP properties is not directly detected, other knots are LyC emitters. Many GPs, like Haro~11, are mergers hosting multiple star-forming knots. So, while GP properties and signatures of radiation-dominated feedback are alone insufficient to predict LyC emission , the multiplicity of starbursts may be important. \citet{Micheva2018} noted the potential role of two-stage starbursts in LyC emitters. On the other hand, Knot A shows clear signatures of density-bounded conditions in the plane of the sky \citep{Keenan2017}. It is therefore likely that Knot~A is leaking LyC transverse to our line of sight, as seen in, e.g., NGC 5253 \citep{Zastrow2011}. This implies that LyC escape must be highly anisotropic, as predicted by simulations \citep{Cen2015} and inferred from observations \citep{ Nestor2011, Flury2022b}. 

Lastly, we note the intriguing coincidence
that the two LyC-leaking knots are the hosts of the only two ULXs in Haro 11 \citep{Prestwich2015, Gross2021}. 
Neither of the ULXs appear to contribute significantly to the LyC emission, especially considering that the stellar populations dominate the UV light by $1-2$ orders of magnitude.
Nevertheless, the X-ray sources may promote LyC escape through accretion-dominated mechanical feedback, where powerful disk winds and jets may clear optically thin channels. A multitude of studies show that reionization-era analogs, such as LBAs and BCDs, have an overabundance of ULXs \citep[e.g.,][]{Dittenber2020, Brorby2014, BasuZych2013, Kaaret2011}. 
Further investigation is thus required into both the ultraviolet emission and mechanical feedback of ULXs, in order to determine their role in LyC escape.

\begin{acknowledgements}
    We thank Paul Draghis and Daniel Schaerer for useful discussions. This work was supported by NASA grant to M.S.O HST-GO-16260.  A.~Adamo acknowledges the support of the Swedish Research Council, Vetenskapsr\aa{}det grant 2021-05559. JMMH is funded by Spanish MCIN/AEI/10.13039/501100011033 grant PID2019-107061GB-C61. D.K. is supported by the Centre National d’\'Etudes Spatiales (CNES)/Centre National de la Recherche Scientifique (CNRS); convention no 230400. The Cosmic Dawn Center (DAWN) is funded by the Danish National Research Foundation under grant DNRF140. Lastly, we thank the anonymous referee for helpful suggestions that improved the manuscript.
        
    This research is based on observations made with the NASA/ESA Hubble Space Telescope obtained from the Space Telescope Science Institute, which is operated by the Association of Universities for Research in Astronomy, Inc., under NASA contract NAS 5–26555. These observations are associated with programs 16260, 15352, and 13017. This research is also based on observations collected at the European Southern Observatory under ESO programme 096.B-0923(A). 
. 
\end{acknowledgements}
The {\it HST} data used in this paper can be found in MAST: \dataset[10.17909/p40f-yy93]{http://dx.doi.org/10.17909/p40f-yy93}
\facilities{HST(COS), VLT(MUSE)}
\software{Starburst99 \citep{Leitherer1999, Leitherer2014}, Astropy \citep{astropy:2022}, \textsc{lmfit} \citep{Newville2016}, Matplotlib \citep{Hunter:2007}}

\bibliography{main}{}
\bibliographystyle{aasjournal}
\end{document}